\documentclass[aps,pra,twocolumn,showpacs,reprint]{revtex4-1}
\usepackage{graphicx}
\usepackage{bm}
\usepackage{amsthm,amsfonts,amstext,amsmath,amssymb}
\usepackage{latexsym}
\usepackage[margin,index]{fixme}
\usepackage[utf8]{inputenc}
\usepackage[makeroom]{cancel}
\usepackage[T1]{fontenc}
\usepackage[usenames,dvipsnames]{color}
\usepackage{bbm}
\usepackage{braket}

\newcommand{\be}{\begin{equation}}
	\newcommand{\ee}{\end{equation}}
\newcommand{\bea}{\begin{eqnarray}}
	\newcommand{\eea}{\end{eqnarray}}

\def\ie{{i.e.},\ }

\newcommand*{\ketbra}[2]{\ensuremath{\ket{#1}\bra{#2}}}
\newcommand*{\com}[2]{\ensuremath{\left[#1, #2\right]}}

\begin{document}
	
	\title{Coupled Atomic Wires in a Synthetic Magnetic Field}
	
	\author{J. C. Budich${}^{1}$, A. Elben${}^{2,3}$, M.~\L\k{a}cki${}^{2,3}$, A.~Sterdyniak${}^{4}$, M. A.~Baranov${}^{2,3}$, P.~Zoller${}^{2,3,4}$}
	\affiliation{${}^{(1)}$ Department of Physics, University of Gothenburg, SE 412 96 Gothenburg, Sweden}
	\affiliation{${}^{(2)}$ Institute for Theoretical Physics, University of Innsbruck, A-6020 Innsbruck, Austria}
	\affiliation{${}^{(3)}$ Institute for Quantum Optics and Quantum Information of the Austrian Academy of Sciences, A-6020 Innsbruck, Austria}
	\affiliation{${}^{(4)}$ Max-Planck-Institute of Quantum Optics, Hans-Kopfermann-Str. 1, D-85748 Garching, Germany}
	\date{\today}
	\begin{abstract}
We propose and study systems of coupled atomic wires in a perpendicular synthetic magnetic field as a platform to realize exotic phases of quantum matter. This includes (fractional) quantum Hall states in arrays of many wires inspired by the pioneering work [Kane et al. PRL {\bf{88}}, 036401 (2002)], as well as Meissner phases and Vortex phases in double-wires. With one continuous and one discrete spatial dimension, the proposed setup naturally complements recently realized discrete counterparts, i.e.~the Harper-Hofstadter model and the two leg flux ladder, respectively. We present both an in-depth theoretical study and a detailed experimental proposal to make the unique properties of the {\em semi-continuous Harper-Hofstadter model} accessible with cold atom experiments. For the minimal setup of a double-wire, we explore how a sub-wavelength spacing of the wires can be implemented. This construction increases the relevant energy scales by at least an order of magnitude compared to ordinary optical lattices, thus rendering subtle many-body phenomena such as Lifshitz transitions in Fermi gases observable in an experimentally realistic parameter regime. For arrays of many wires, we discuss the emergence of Chern bands with readily tunable flatness of the dispersion and show how fractional quantum Hall states can be stabilized in such systems. Using for the creation of optical potentials Laguerre-Gauss beams that carry orbital angular momentum, we detail how the coupled atomic wire setups can be realized in non-planar geometries such as cylinders, discs, and tori.	
\end{abstract}
\maketitle

\section{Introduction}

Synthetic matter based on ultracold atomic gases in optical
potentials \cite{Bloch2008} has established itself as one of the most promising experimental routes to realize exotic quantum matter \cite{Lewenstein2012,NathanReview}.
Recent theoretical \cite{Jaksch2003,Duan2006,GerbierDalibard,DalibardReview,Cooper2011,Cooper2013,GoldmanReview}
and seminal experimental \cite{Aidelsburger2011,Sengstock,Ketterle2013,Aidelsburger2013,Atala2014,Aidelsburger2015}
results on the implementation of synthetic magnetic fields and even
non-Abelian gauge fields such as synthetic spin-orbit coupling \cite{Galitski2013} have
paved the way towards realizing topological insulators \cite{HasanKane2010,QiZhang2011}
with ultracold atoms \cite{NathanReview}. In particular, the Harper Hofstadter
(HH) Hamiltonian \cite{Hofstadter1976}, introduced to describe the
motion of electrons in a 2D square lattice exposed to a perpendicular
magnetic field has recently been experimentally realized with cold
atoms in optical lattices \cite{Aidelsburger2013,Ketterle2013}.

\begin{figure}[t!]
\includegraphics[width=86mm]{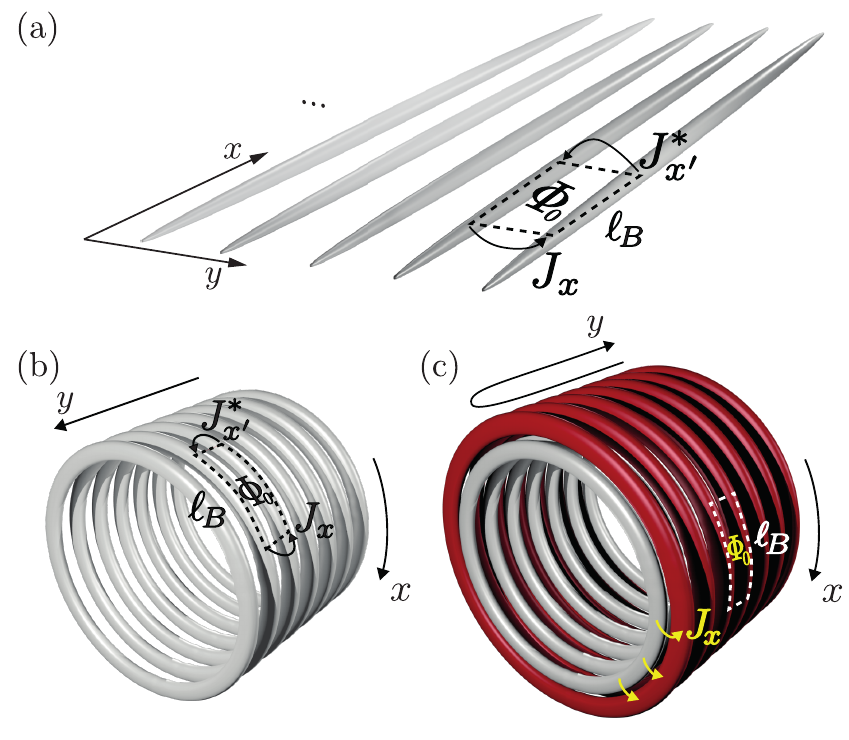} \caption{(Color online) Systems of wires in various geometries coupled by Raman-induced
hopping with amplitude $J_{x}=Je^{i\phi x}$ inducing artificial magnetic
field with a flux quantum $\Phi_0$ threading a unit cell of length
$l_{B}$ along the wires. Panel (a) planar geometry. Panel (b) structure
of a cylinder spanned by a set of parallel rings. Panel (c) geometry
of a torus constructed using two nested coaxial cylinders coupled only at the  end rings of each cylinder via  Raman-induced hopping.}
\label{fig:setup} 
\end{figure}

In the present work, we propose and present a detailed study of ultracold atomic gases confined to an array
of coupled wires forming a 2D system with \emph{one continuous and
one discrete dimension} with synthetic magnetic fields.
Employing Raman-assisted tunneling techniques \cite{Jaksch2003,GerbierDalibard,Cooper2011}
to generate artificial magnetic fluxes, we introduce an experimentally
feasible scheme for the implementation of a semi-continuous generalization
of the HH model. We are interested in particular in a physical realization of this model for various geometries: this includes a planar setup (see Fig.\:\ref{fig:setup}(a)), which can be realized as a variant of existing cold atom experiments with optical lattices,  or as disks, cylinders,
or tori (see Fig.\:\ref{fig:setup} (b)-(c)), which is achieved by employing optical potentials created from  Laguerre-Gauss laser beams carrying orbital angular momentum
(OAM) \cite{Allen1992, Allen2003, Padget2008}. 
Below, we will consider
both bosonic and fermionic atomic quantum many-body systems, and we will explore the
rich variety of exotic phases of matter that can be realized in such
coupled atomic wires.

The non-interacting part $H_{0}$ of the Hamiltonian describing an
array of coupled wires is given by 
\begin{align}
H_{0}=\sum_{y}\int\text{d}x\left[\psi_{x,y}^{\dag}\frac{p_{x}^{2}}{2m}\psi_{x,y}+(J_{x}\:\psi_{x,y}^{\dag}\psi_{x,y+a}+\text{h.c.})\right].\label{eqn:ham1}
\end{align}
Here, $\psi_{x,y}$ annihilates a particle at position $\mathbf{r}=(x,y)$,
and $J_{x}=J\text{e}^{i\phi x}$ ($J>0$) denotes the complex hopping
amplitude in the discrete $y$-direction for atoms of mass $m$, where $\phi$ determines
the homogeneous synthetic magnetic flux of the system, defining a magnetic length $l_{B}\equiv 2\pi/\phi$, and $a$ denotes
the lattice spacing in $y$-direction. We note that for non-planar
geometries (see Fig.\:\ref{fig:setup} (b)-(c)), the coordinates
$x,y$ should be understood as local coordinates tangent to the curved
2D system, where $x$ always represents the continuous (i.e.~for coupled
rings circumferential) direction, and $y$ refers to sites in the
discrete direction. In the atomic context, $H_{0}$ can be realized
both with fermionic and bosonic atoms. The long-wavelength scattering
between the atoms gives rise to a (two-body) interaction $H_{I}$,
and the total Hamiltonian is hence given by 
\begin{align}
H=H_{0}+H_{I}.\label{eqn:hamint}
\end{align}
As the relevant scattering channels differ for bosons and fermions
thus giving rise to different forms of $H_{I}$, we will specify $H_{I}$
for the individual examples considered in this work.

In a condensed matter context, similar models have been introduced by Kane et al.~\cite{KaneCoupledWires2002,TeoKane2014}
for electronic systems to study fractional quantum Hall (FQH) physics
\cite{Tsui1982,Laughlin1983,Prange1990} in tunnel-coupled nano-wires,
known as the coupled wire construction.
There, the relevance in the renormalization group sense of specific tunnel-coupling terms occurring as
perturbations to the strongly correlated uncoupled 1D wires is investigated.
Here, we focus on a different limit, where the strongly tunnel-coupled
wires form quasi-flat partially filled Chern bands \cite{Bauer2015}
which, adding interactions, host FQH states. Below, we argue how FQH
states such as Laughlin states \cite{Laughlin1983} can be stabilized
in this setup and present a numerical analysis for realistic parameters
in support of these claims. Remarkably, here the flatness of the lowest
band can be directly tuned by changing the hopping strength $J$.
While the controlled realization of coupled wire systems in conventional
materials is challenging, the remarkable flexibility and tunability
of ultracold atomic gases may open a new playground for the synthetic
realization of topological phases of quantum matter.

For a minimal system of {\emph{two coupled wires}}, we compare for
both bosonic and fermionic gases the phase diagram to that of a two-leg
ladder – the well studied discrete analog \cite{Orignac2001,Atala2014,Hugel2014,Mancini2015,Piraud2015,Wei2014}. From an experimental point of view, a key challenge is a sufficiently strong couplings $J$ (with Raman schemes to realize the artificial gauge field)  in relation to the temperature. We will address this problem by proposing a setup of a {\em double wire with subwavelength separation} extending recent work \cite{Zoller2016} (see also Ref.\:\cite{Gorshkov2016}), which allows a strong enhancement by Raman-induced tunneling orders of magnitude of the relevant couplings over standard optical lattices schemes with spacing $a =\lambda/2$ , where $\lambda$ is the wavelength of the light. In the case of a weakly or
non-interacting gas of fermionic atoms, subwavelength double wires promise observation of chiral
edge currents with non-analytic behavior for atoms at Lifshitz transitions
of the Fermi surface.

The remainder of this paper is structured as follows. In Section
\ref{sec:Models}, we theoretically analyze the model of coupled wires,
focusing on geometries and parameter regimes that are of immediate
relevance for the proposed implementation with ultracold atomic gases
that is detailed in Section \ref{sec:Implementation}. Finally, in
Section \ref{sec:Conclusions} we provide a concluding discussion
of our results.

\section{Chern bands and topological phases in coupled atomic wires}

\label{sec:Models} In this Section, we discuss the properties of
systems described by the discrete-continuous Hamiltonian with artificial
gauge fields given in Eqs.~\eqref{eqn:ham1} and \eqref{eqn:hamint}
in various geometries such as planar, cylindrical and toroidal, the
experimental implementation of which is discussed in detail in the
subsequent Sec.~\ref{sec:Implementation}. In anticipation of the
atomic physics analysis, we will identify the relevant physical parameters and energy scales, to be compared with the experimentally achievable parameters for various setups in Sec.~\ref{sec:Implementation}.
 In Sec.~\:\ref{sec:Models:torus},
we discuss the topological band structure of the Hamiltonian \eqref{eqn:ham1}
in the presence of translation invariance, i.e.~in the geometry of
an infinite plane or a (finite) torus. In Sec.~\ref{sec:Models:cylinder},
we turn to a cylindric or planar ribbon geometry of $N$ wires, and
investigate the occurrence of edge states. In Sec.~\ref{sec:Models:twoWires},
a minimal, analytically treatable case of $N=2$ parallel wires or
rings is analyzed. Adding interactions, we study in Sec.~\ref{sec:Models:FQH}
a strongly correlated bosonic gas. In the framework of exact diagonalization,
we discuss the realization of robust fractional Quantum Hall physics
in flat-band systems.

\subsection{Translation invariant model}

\label{sec:Models:torus}

\begin{figure}[b]
\includegraphics[width=86mm]{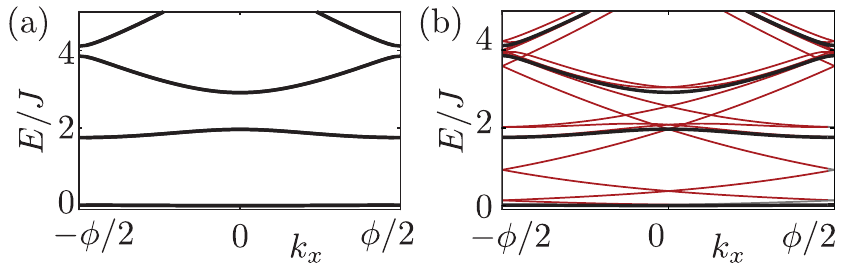} \caption{(Color online) Band structure $k_{x}\in[-\phi/2,-\phi/2)$ ($\phi/2\equiv \pi/\ell_B$) of the three lowest bands of the free model Hamiltonian
\eqref{eqn:hky}. Panel (a) shows a closed geometry of a torus with $N\gg 1$ wires, or
for a planar system with an infinite number of wires. Panel (b) shows
energy levels in the limit of $N\gg 1$ of parallel wires; black lines
denote energy levels of states supported in the bulk of the system, red lines show edge states. In both plots, we have chosen $\tilde  J\equiv J/(\hbar^{2}\phi^{2}/2m)=0.8$.}
\label{fig:BandsBerry} 
\end{figure}

We start by considering the free Hamiltonian in Eq.~\eqref{eqn:ham1}.
Assuming translation invariance in $y$-direction, we can separate
the corresponding motion by using the plane waves with a quasimomentum
$k_{y}\in[-\pi/a,\pi/a]$ with $a$ the lattice spacing in the $y$-direction. For a given $k_{y}$, the first quantized
Hamiltonian in the $x$-direction takes the form 
\begin{align}
h_{x,k_{y}}=-\frac{\hbar^{2}}{2m}\frac{d^{2}}{dx^{2}}+2J\cos(\phi x+k_{y}a).\label{eqn:hky}
\end{align}
Eq.\:\eqref{eqn:hky} describes a 1D particle moving in a potential
with a period given by the magnetic length $l_{B}=2\pi/\phi$. The corresponding energy scale is given by the   magnetic (recoil) energy $E_{R}^{M}\equiv \hbar^{2}\phi^{2}/2m$. The
form of Eq.\:\eqref{eqn:hky} parametrically depends on $k_{y}$
in a way that makes the Hall response of the system intuitively clear.
Varying $k_{y}$ by $2\pi/a$, i.e.\ once around the Brillouin zone
(BZ), the potential slides $\ell_B = 2\pi/\phi$ in $x$-direction, i.e.\ by
one period of the potential in Eq.\:\eqref{eqn:hky}. In agreement
with this intuitive picture, the Chern number $\mathcal{C}^{\alpha}$
of a Bloch band $\alpha$ \cite{TKNN1982}, defined in terms of the projection $P_{k}^{\alpha}$
onto the Bloch state $\lvert u_{k}^{\alpha}\rangle$ at lattice momentum
$k=(k_{x},k_{y})$ as 
\begin{align}
\mathcal{C}^{\alpha}=\frac{i}{2\pi}\int_{BZ}\text{d}^{2}k\:\text{Tr}\left\{ P_{k}^{\alpha}\left[(\partial_{k_{x}}P_{k}^{\alpha}),(\partial_{k_{y}}P_{k}^{\alpha})\right]\right\} =1\label{eqn:Chern}
\end{align}
is always one, irrespective of the parameter values as long as the
band is separated by an energy gap from the other bands. As is clear
from the translational symmetry of Eq.\:\eqref{eqn:hky}, the spectrum
of the system is independent of $k_{y}$ and so is the concomitant
Berry curvature $\mathcal{F}_{k}^{\alpha}=i\text{Tr}\left\{ P_{k}^{\alpha}\left[(\partial_{k_{x}}P_{k}^{\alpha}),(\partial_{k_{y}}P_{k}^{\alpha})\right]\right\} $.

An example of the $k_{x}$-dependent band structure
is shown in Fig.\:\ref{fig:BandsBerry}. Interestingly, and in contrast
to the discrete HH model, increasing the hopping strength $J$ here
provides a simple knob to make the energy spectrum flat as soon as
the dimensionless parameter $\tilde J \equiv J/E_R^M\gtrsim1$. The flatness is characterized
by the ratio of the gap to the first excited state to the width of
the lowest band. For the specific parameters in Fig.\:\ref{fig:BandsBerry}
we achieve a flatness ratio of 137 already for $\tilde J =0.8$
(for scaling of the flatness, and of the Berry curvature we refer
to Appendix \ref{app:BandFlatness}).

\subsection{Finite number of wires}

\label{sec:Models:cylinder}

We now consider the case of open boundary conditions in the discrete
direction, reflecting a finite number $N$ of infinitely long wires
in a planar geometry or $N$ parallel rings in a cylindrical geometry.
In this case, $k_{y}$ is not a good quantum number. However, each
individual wire or ring still possesses $k_{x}$-eigenstates $|y,k_{x}\rangle$,
labeled by $y=y_{1},\ldots,y_{N}.$ The eigenvectors of the Hamiltonian
\eqref{eqn:ham1} are then of the form 
\begin{align}
|\psi\rangle=\sum_{j=1}^{N}c_{j}|y_{j},{k}_{x}+\phi(j-j_{0})\rangle,\,k_{x}\in[-\phi/2,\phi/2)\label{eqn:singleParticleWavefunction}
\end{align}
for some $j_{0}\in\mathbb{Z}.$ For the eigenvectors where the $c_{j}$'s
are supported mostly in the bulk of the system, the eigenenergies
are similar (for $N\gg1$) to those for an infinite case. In particular,
they are independent of the actual position of the wavefunction in
the bulk and feature the same band flatness. Reflecting the unit Chern
number of the bands, there are also edge states. For them, the $|c_{j}|$
are localized on a few ($\sim\sqrt[4]{\tilde J}$)
wires, or rings at the boundary and their energies cross the bulk gap
between different bands.

Probing such edge modes is possible by inducing a Hall response to
a time-dependent magnetic flux along the lines of the Laughlin argument
\cite{Laughlin1981}. To that end, in a many-particle system of non-interacting
ultracold fermions a synthetic magnetic probe field $\mathbf{B}=B_{0}{\mathbf{e}_{y}}$
is induced by increasing, in a time-dependent manner, the velocity
of the gas in the $x$ direction. This can be done either by direct
stirring \cite{Philips2013} or transfer of optical momentum \cite{Philips2007}.
The result is a measurable transfer of population between edge-modes
\cite{Zoller2016a}, which allows to verify the topological nature
of the Hamiltonian \eqref{eqn:ham1}.

\subsection{Minimal setup: Two coupled wires}

\label{sec:Models:twoWires}

In this Section, we analyze the minimal setup consisting of only $N=2$
coupled wires. It represents the simplest system showing non-trivial
artificial magnetic field effects and is readily realized experimentally (see Sec.~\ref{sec:Implementation})
\cite{Schmiedmayer2006,Schmiedmayer2006b,Atala2014}. The physics
of many particles in the discrete analog of our system, the two-leg
ladder, has been studied intensively theoretically and experimentally
\cite{Orignac2001,Atala2014,Hugel2014,Mancini2015,Piraud2015,Wei2014}.
For bosonic particles, the two-leg ladder mimics to some extent the behavior of
a Type-II superconductor, exhibiting Meissner and vortex phases in
a weak and strong external magnetic field, respectively \cite{Orignac2001,Hugel2014,Piraud2015}.
These phases have been demonstrated experimentally \cite{Atala2014}.
Here, we find similar phenomena in the bosonic continuous double wire.
For fermions, by contrast, we show that no vortex phase is present.
Instead, weak and strong field regimes are distinguished by the presence
or absence of Lifshitz transitions, respectively, manifesting themselves
in non-analyticities of the chiral current. 

In the absence of interactions, the physics of the double wire system
is governed by the Hamiltonian \eqref{eqn:ham1} where the sum in
the discrete direction $y$ is reduced to two wires, labeled with
$L$ (left) and $R$ (right), respectively. The resulting Hamiltonian
\eqref{eqn:ham1} can be easily diagonalized in Fourier space for
the continuous $x$-direction, exhibiting two energy bands 
\begin{align}
E_{\pm}(k_{x})=\frac{\hbar^{2}k_{x}^{2}}{2m}+\frac{\hbar^{2}\phi^{2}}{2m}\left(\frac{1}{4}\pm\sqrt{\frac{k_{x}^{2}}{\phi^{2}}+\tilde{J}^{2}}\right)\;.
\end{align}
Here, $\tilde{J}=J/E_R^M$ is the dimensionless interwire
coupling introduced above. The   wave functions are given by
\begin{align}
\Psi_{k_{x}}^{(\pm)}(x)
=\left[2\sqrt{1+\left(\frac{k_{x}}{\tilde{J}\phi}\right)^2}\right]^{-\frac{1}{2}}\!\!\begin{pmatrix}f^{\pm\frac{1}{2}}\;e^{-i\frac{\phi}{2} x}\\
\pm f^{\mp\frac{1}{2}}\;e^{i\frac{\phi}{2} x}
\end{pmatrix}e^{ik_{x}x}\label{eq:wavefunctions}
\end{align}
with $f=\sqrt{1+(k_{x}/\tilde{J}\phi)^{2}}-(k_{x}/\tilde{J}\phi)$,
where upper (lower) component corresponds to the right ${\Psi}_{R}(x)$
{[}left ${\Psi}_{L}(x)${]} wire. 

\begin{figure}[t]
\includegraphics[width=86mm]{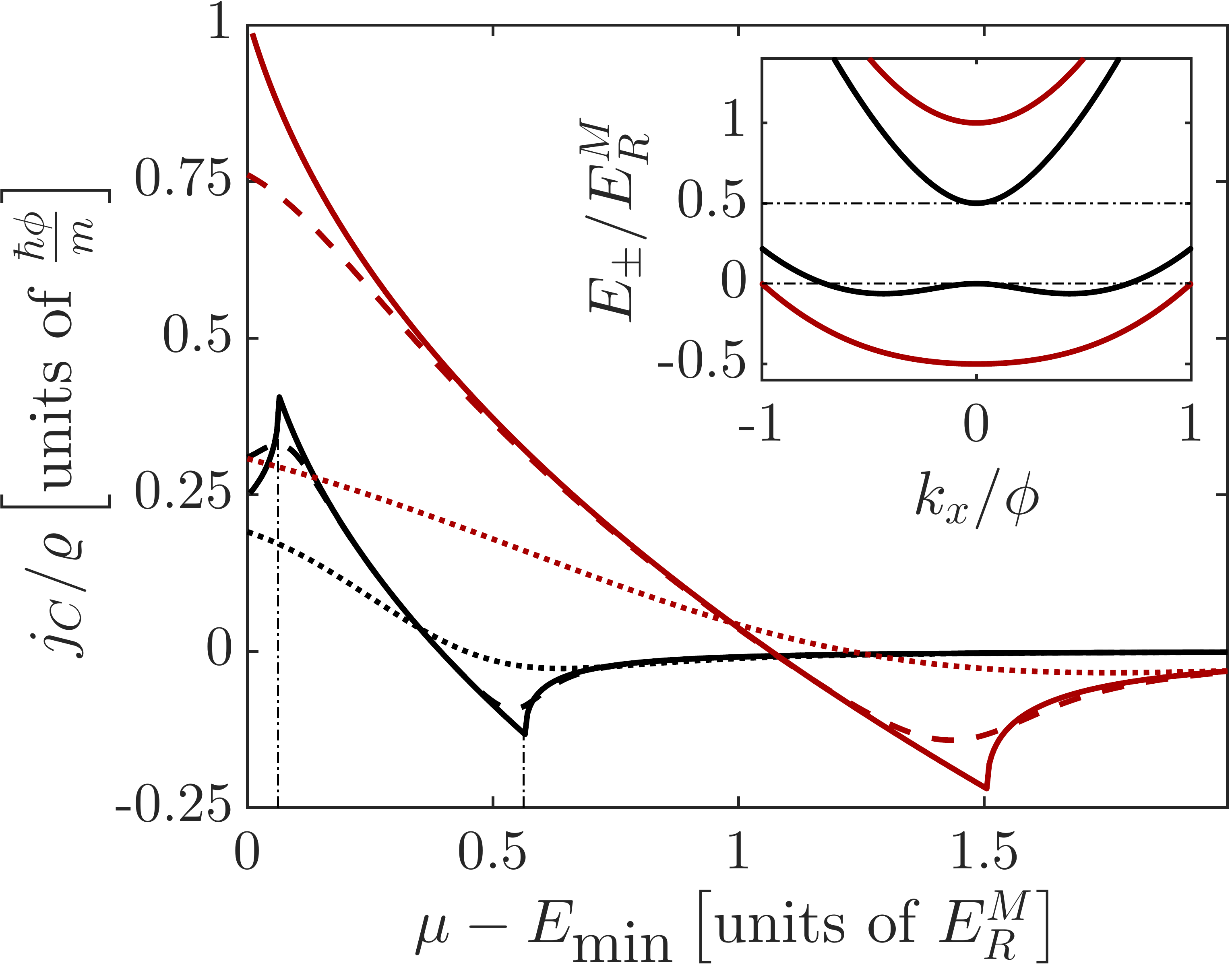}
\caption{(Color online) Chiral current $j_{C}$ for non-interacting fermions normalized by
the particle density $\varrho$ as a function of the chemical potential
$\mu$ of the lowest band for different values of the rescaled coupling $\tilde{J}$,
black: $\tilde{J}=0.25$, red: $\tilde{J}=0.75$ and for different
ratios of temperature and coupling $k_{B}T/J=0$ (solid), $k_{B}T/J=0.1$
(dashed) and $k_{B}T/J=0.5$ (dotted). $E_{\text{min}}$ denotes the respective minimum energy of the lowest band.  Inset: Corresponding band structure.
The upper band for the coupling $\tilde{J}=0.75$ (red) has
much higher energy (not visible in the inset).}
\label{fig:currentfermion2} 
\end{figure}

The upper band $E_{+}(k_{x})$ has in the entire parameter regime
a single global minimum at $k_{x}=0$. The form of the lower band
$E_{-}(k_{x})$ depends on $\tilde{J}$. For $\tilde{J}<1/2$, it
exhibits two degenerate minima at $k_{x}^{(1,2)}=\pm k_{g}$ with
$k_{g}=(\phi/2)\sqrt{1-4\tilde{J}^{2}}$ and corresponding eigenstates
$\ket{\pm k_{g}}=\ket{\pm k_{g}}_{-}$. From Eq. (\ref{eq:wavefunctions})
for their wave functions $\Psi_{\pm k_{g}}^{(-)}(x)$ we see that
the state $\ket{+k_{g}}$ is mostly located on the right wire, while
the state $\ket{-k_{g}}$ on the left one. For $\tilde{J}\geq1/2$,
the two minima merge into a single global minimum, located at $k_{x}^{(0)}=0$
(see inset of Fig.\:\ref{fig:currentfermion2}).

Non-interacting bosons at zero temperature occupy for $\tilde{J}<1/2$
an equally weighted superposition of the two lowest-energy states $\ket{\pm k_{g}}$
[we assume the same number of particles in each wire, see Eq.~\eqref{eq:wavefunctions}{]}.
Hence, the ground state for total particle number $M$ is given by
$\ket{G}=2^{-M/2}\left(\ket{k_{g}}+e^{i\theta}\ket{-k_{g}}\right)^{\otimes M}$
with an arbitrary phase $\theta$. The difference $2k_{g}$ of the
momentum components of this state introduces a length scale in the
$x$-direction, leading to a periodic spatial modulation of the particle
density 
\begin{align}
\varrho(x) & =\langle{\Psi}_{L}^{\dagger}(x){\Psi}_{L}(x)\rangle=\langle{\Psi}_{R}^{\dagger}(x){\Psi}_{R}(x)\rangle\nonumber \\
 & \sim\cos(2k_{g}x-\theta)+\mathrm{\text{const}},
\end{align}
and of the chiral current
\[
j_{C}(x)=j_{L}(x)-j_{R}(x)\sim\tilde{J}\cos(2k_{g}x-\theta)+2\tilde{J}^{2},
\]
where $j_{L(R)}$ denotes the probability current in left (right)
wire. This represents the vortex phase. On the contrary, for $\tilde{J}\geq1/2$,
we find a Meissner phase with constant particle densities in both
wires, and the constant chiral current $j_{C}$ reaching its maximum.

Non-interacting fermions at zero temperature occupy all states with
energies below the chemical potential $\mu$, and their density $\varrho$
is always constant such that no vortex phase is present. However,
the chiral current, depending on the chemical potential $\mu$, allows
to distinguish the fermionic analog of the two bosonic phases. In
Fig.~\ref{fig:currentfermion2}, the chiral current is shown as
a function of $\mu$, normalized by the particle density $\varrho$
(solid lines for zero temperature). We observe a non-analytic behavior
of the current at values of $\mu$ where the Fermi surface changes
its topology, i.e.\ the system undergoes a Lifshitz transition \cite{Lif1960}.
For $\tilde{J}<1/2$, which for particles with bosonic statistics
would correspond to a vortex phase, two such Lifshitz transitions
occur. The first one, which is only present for $\tilde{J}<1/2$,
takes place when the chemical potential touches the maximum of the
lower band. The second one, existing for any non-vanishing coupling
$\tilde{J}$, occurs when the chemical potential touches the bottom
of the upper band. Hence, the presence and absence of the first Lifshitz
transition determined by the form of the lowest band represents the
fermionic analog to the bosonic vortex and Meissner phase, respectively.

At finite temperatures the non-analyticities in the chiral current
are smeared out, but still visible for temperatures much smaller than
the band gap (see dashed lines in Fig.\:\ref{fig:currentfermion2}).
To resolve the non-analyticities in the chiral current in an experiment,
the temperature has to be sufficiently low compared to energy scales
set by coupling $J$ and flux $\phi$. For the double wire with subwavelength separation discussed in Sec.~\ref{sec:Implementation:wires:construction}, these non-analyticities should be observable with current experimental technology (dashed lines in Fig.\:\ref{fig:currentfermion2}). On the contrary, for smaller couplings strengths, typical for standard setups with wire
separation $\lambda/2$, the chiral current is featureless. For details on the achievable coupling parameters, we refer to Sections\:\ref{sec:Implementation:OL} and \ref{sec:Implementation:wires:tunneling} \cite{note1}.

Finally, we note that the phase distinction for both, the fermionic
and the bosonic case, is determined by low momentum properties of
the bands. Hence, it can be found likewise in the continuous double
wire and in the discrete two-leg ladder.

\subsection{Fractional quantum Hall states in coupled atomic wires}
\label{sec:Models:FQH}

The FQH effect is the paradigmatic example of strongly correlated phases of matter that can emerge in topological bands. It typically occurs when a band with a non-zero Chern number is partially filled by strongly interacting particles. For certain rational filling factors of the band, interactions lead to incompressible states that can host anyonic fractionally charged excitations. The typical hierarchy of energy scales where this occurs is characterized by an interaction strength that is much larger than the band width but much smaller than the band gap, hence the interest in flat bands. While this phenomenon can occur both for fermionic and bosonic particles, the filling factor at which FQH states appear depends on the statistics of the particles. For example, the most prominent FQH state for fermions appears at filling $1/3$ whereas its bosonic counterpart emerges at filling $1/2$. While the FQH effect was experimentally discovered in partially filled Landau levels of a 2D electron gas exposed to a strong perpendicular magnetic field, \ie for electrons, many theoretical studies showed its realization in the context of cold atomic gases, both for bosons and fermions, in various setups such as rotating traps~\cite{Cooper2001,Regnault2004,Baranov2005}, optical lattices with an artificial gauge field~\cite{Lukin2005,Lukin2007}, optical flux lattices~\cite{Cooper2013,Sterdyniak2015}, and in Chern insulators~\cite{Neupert2011,Sheng2011,Regnault2011} (for reviews see \cite{Cooper2008,Bergholtz2013,Parameswaran2013}).

Below, we investigate the realization of bosonic FQH states such as Laughlin states \cite{Laughlin1983} in an array of coupled wires where the non-interacting part of the Hamiltonian is given by Eq.~\eqref{eqn:ham1}. The natural interaction between the bosonic atoms, microscopically stemming from $s$-wave scattering, is an on-site interaction in the discrete $y$-direction and a contact-interaction in the $x$ direction. In first quantization, $H_I$ is hence given by 
\begin{equation}
H_{I} = U \sum_{i<j} \delta(x_i-x_j)\delta_{y_i,y_j} \label{ref:Vint}.
\end{equation}
Physically, this model is in between the HH model for bosons with on-site interactions and a model of bosons with contact interactions in a rotating trap, which both were shown to host the $\nu=1/2$ Laughlin state in Ref.~\cite{Lukin2005,Lukin2007} and Ref. \cite{Cooper2001}, respectively. Intuitively, we hence also expect this to be the case in the present model.

Interestingly, the present model is also from a formal analytical perspective in between the lowest Landau level (LLL) in the continuum and the HH on a lattice. Relating the lattice to the continuum problem, a modified HH model with longer ranged hopping known as the Kapit-Mueller model has been constructed in Ref. \cite{KapitMueller}. It features an exactly flat lowest band spanned by the same single-particle states as the lowest Landau level, simply restricted to the discrete positions of the lattice sites.
Then, as the bosonic $1/2$ Laughlin state is the exact ground state of the half filled LLL annihilating the contact interactions, its restriction to the discrete lattice is the exact ground state of the bosonic Kapit-Mueller model with on-site interactions \cite{KapitMueller}. For the present semi-continuous HH model, the lowest band is exactly flat in the limit $J\rightarrow \infty$, i.e. without introducing longer range hoppings, and the single particle states that span it are those of the LLL, restricted to the semi-continuous geometry of the coupled wires. As $H_I$ in Eq. (\ref{ref:Vint}) is simply the contact interaction from the continuum restricted to the coupled wires, the concomitant restriction of the $1/2$ Laughlin state from the LLL annihilates these interactions and thus must become the ground state of our present model for large $J$.

For finite systems and with the experimentally realistic parameters $J=0.05 E_R, l_B=4 \lambda$, the occurrence of a Laughlin state is hard to assess analytically. In the remainder of this section, using exact diagonalization, we hence numerically investigate this situation, focusing on the emergence of the $\nu=1/2$ Laughlin state in a system with periodic boundary conditions. We provide strong evidence for the occurrence of this state using both energy and entanglement spectroscopy, and give a physical picture for the resulting phase diagram of the model as a function of $J$. As an additional probe for the robustness of the Laughlin phase, we carefully checked that the groundstate manifold is stable under flux insertion in both directions.

\subsubsection{Energy spectroscopy}

As commonly done in the study of FQH systems both in the continuum and on a lattice~\cite{Regnault2011}, we use the so called flat band approximation: we project the interacting Hamiltonian to the lowest band of the single-particle model, neglecting the admixture of all other bands as well as the band dispersion. This is a valid approximation as long as the interaction strength is much larger compared to the band width and much smaller than the band gap. We consider $N_B$ bosons in a system of $N_y$ coupled wires of length $N_xl_B$, so that the system is pierced by $N_xN_y$ magnetic flux quanta and the filling of the lowest band is given by $\nu=\frac{N_B}{N_xN_y}$. This allows us to diagonalize simultaneously the Hamiltonian and the total momentum operator in both directions, defining thus two quantum numbers $(K_x,K_y)$ for every state.

\begin{figure}[htb]
	\includegraphics[width=0.98\linewidth]{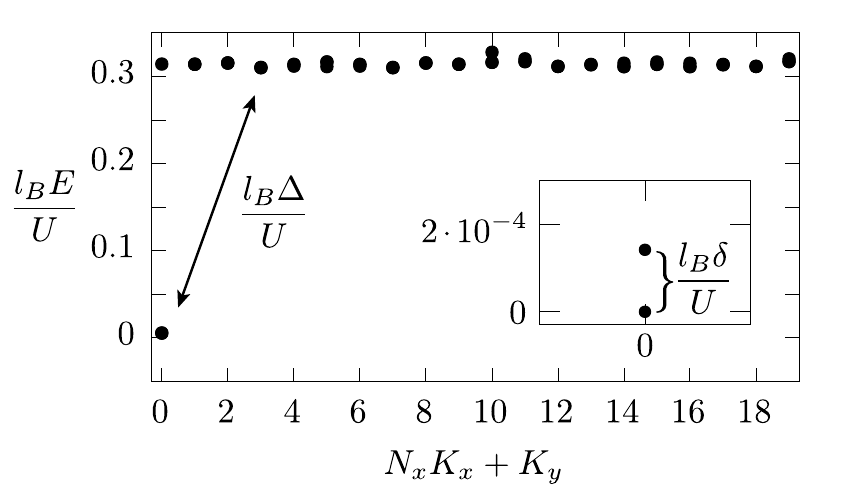}
	\caption{Spectrum of the interaction Hamiltonian~\eqref{ref:Vint} for $N_B=10$, $N_x=2$, $N_y=10$, $l_B=4 \lambda$  and $J=0.05E_R$. The energy levels are plotted with respect to a linearized momentum: $N_xK_x+K_y$. The inset shows a zoom on the ground state manifold which is quasi two-fold degenerate as expected for the $\nu=1/2$ Laughlin state. The energy splitting $\delta$ between the ground states and the energy gap $\Delta$ are defined graphically in the figure.}
	\label{fig:energy_spec}
\end{figure}		

We focus on the $\nu=1/2$ filling where the bosonic Laughlin state is a natural candidate for describing the ground state. On the torus, it exhibits a characteristic two-fold degeneracy due to many-body magnetic translation invariance~\cite{Haldane1985}. However, this degeneracy is lifted in finite lattice systems, such as the one considered here and the energy splitting $\delta$ between the two lowest energy states should be much smaller than the energy gap $\Delta$ to the third lowest energy state. A typical energy spectrum for the parameters identified in Section \ref{sec:Implementation} is shown in Fig.~\ref{fig:energy_spec} for $N_B=10$, $N_x=2$ and $N_y=10$. It shows the expected quasi two-fold degenerate ground state manifold (with an energy splitting $\delta= 1.4~10^{-4}~U/l_B$) separated by a gap $\Delta= 0.3~U/l_B$ from higher energy states. Depending on the geometry investigated, the ground states appear in different momentum sectors. These momentum sectors can be predicted~\cite{Regnault2011} and we checked that our numerical results match these predictions.

\begin{figure}[htb]
	\includegraphics[width=0.98\linewidth]{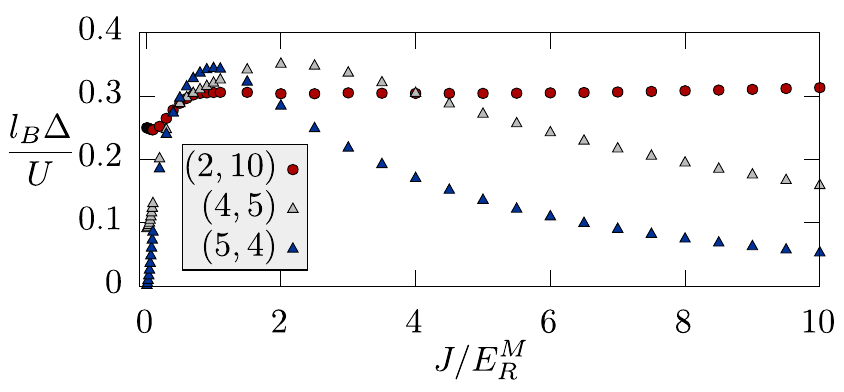}
	\caption{(Color online) Energy gap between the second and the third lowest energy states for $N_B = 10$, $\phi=2\pi/a$ and different aspect ratios as a function of the inter-wire hopping. The different system sizes are indicated by the values of $(N_x,N_y)$ in the legend.}
	\label{fig:energy_Gap}
\end{figure}

To investigate more generically the role of interaction in our model, we restrict ourselves to the case $\phi=2\pi/a$ in Eq.~(\ref{eqn:hky}). This can be done without loss of generality since a generic $\phi$ can be mapped onto the $\phi=2\pi/a$ case via a rescaling of the length in the $x$ direction and a change of the recoil energy. For this case, we show the energy gap as a function of the hopping between the wires for $N_B=10$ and different aspect ratios in Fig.~\ref{fig:energy_Gap}. The aspect ratio seems to have an important effect on the stability of the Laughlin phase. Indeed, while for $(N_x = 2,N_y = 10)$ the gap is much more stable with increasingly large $J$, the gap decreases clearly for $(N_x = 4,N_y = 5)$ and even more for $(N_x = 5,N_y = 4)$. As we explain below, the crucial figure in this context is actually not the aspect ratio, but only $N_y$. The ostensible dependence on the aspect ratio is simply due to the fact that, for computational reasons, we cannot increase $N_y$ without decreasing $N_x$ at the same time.  

The $N_y$ dependence in the stability of the FQH phase can be physically understood at an intuitive level when looking at Eq. (\ref{eqn:hky}). With increasing $J$, the localization length of the lowest Wannier function in $x$-direction (at a given $k_y$) decreases as $J^{-(1/4)}$. The number $N_y$ then determines how many such equidistantly spaced orbitals there are per magnetic length $l_B$ in $x$-direction. At small enough $N_y$, these Wannier orbitals have virtually no overlap in $x$-direction, and hence allow the bosons to avoid the contact interaction, even in trivial product states forming a charge density wave (CDW). We hence expect in agreement with Fig.~\ref{fig:energy_Gap} that in finite systems with a given $N_y$, the Laughlin phase breaks down with increasing $J$, at a critical $J_c$ that increases with $N_y$. This picture will be corroborated by the subsequent entanglement spectroscopy. Note that these findings are by no means in contrast to the above analytical arguments stating that for $J\rightarrow \infty$ the ground state in the thermodynamic limit will be the exact Laughlin wave-function, but rather reflect a simple order of limits problem. This is because for $N_y$ going to infinity, the discrete $k_y$ values will become dense, and thus there will be an overlap of neighboring Wannier functions for arbitrary $J$.

\subsubsection{Entanglement spectroscopy}

The two-fold degeneracy of the ground state is only a necessary condition for the system to be in the Laughlin phase. Indeed, a charge density wave (CDW), breaking translational symmetry, would also exhibit the same ground state degeneracy and is a known competing phase. Unlike symmetry breaking phases, topological phases do not possess a local order parameter that could be computed to probe the nature of the phase and are, thus, difficult to identify unambiguously, especially in numerical studies. To solve this issue, building upon the concept of entanglement entropy, Haldane and Li introduced the orbital entanglement spectrum (OES)~\cite{Haldane2008}, i.e. the spectrum of the reduced density matrix of a subsystem defined by a subset of the single particle orbitals. It allows to identify the edge theory of the ground state and, in many cases, the corresponding topological phase. For torus geometries and lattice systems, the particle entanglement spectrum (PES), a concept closely related to the OES, is a very practical tool to identify FQH states~\cite{Sterdyniak2011}. It is defined through a partition of the system in two subgroups $A$ and $B$, made of $N_A$ and $N_B-N_A$ particles. The PES is the spectrum of $H_A = -\log \rho_A$ where $\rho_A$ is the reduced density matrix obtained by tracing out the $B$ particles and $H_A$ is called the entanglement Hamiltonian. As $\rho_A$ commutes with the total momentum operators for particles in the $A$ part, we use these operators quantum numbers $(K_{x,A},K_{y,A})$ to block-diagonalize $\rho_A$. Unlike the OES in which the geometry of the system is altered, the geometry is unchanged by this procedure and the PES allows to access the physics of the bulk excitations. 

For \emph{ideal FQH model states}, such as the Laughlin wave-function, the PES is made of a number of levels equal to the number of quasihole-states corresponding to the system size under investigation. For example, if we consider the $\nu=1/2$ Laughlin state on the torus with $4$ particles and trace over two of them, the PES will reflect the Laughlin quasihole states with two particles and four quasiholes: not only the number of levels in the PES is then equal to $20$, i.e. the number of quasihole-states, but the eigenstates of $H_A$ are precisely quasihole-states as they are zero energy states of the contact interaction. All other eigenstates of the reduced density matrix have zero eigenvalues and are thus associated with levels at infinite entanglement energy in $H_A$. For any given number of particles and quasiholes, the number of quasihole states can be predicted using Haldane's generalized exclusion principle~\cite{Haldane1991} and are thus direct fingerprints of the anyonic nature of the excitations. For the $\nu=1/2$ Laughlin state on the torus, the number of such states with $N$ particles in $N_{\phi}$ flux quanta is given by $\mathcal{N}(N,N_{\phi}) = N_{\phi} \frac{(N+n-1)!}{N!n!}$, where $n= N_{\phi}-2N$ is the number of quasiholes~\cite{Fayyazuddin1996}. In the PES framework, $N$ should be replaced by $N_A$ and $N_{\phi}$ by $2N$ in the previous formula.

\emph{Away from ideal model states} but still in the same FQH phase, the PES generically does not exhibit levels at infinite energy, but still displays a significant gap below which the number of states is still given by the number of quasihole-states. When this is the case, we conclude that the state is in the same topological phase as the model state. Unlike the counting of quasihole-states in the energy spectrum, the entanglement spectrum was shown to be able to distinguish between a CDW and a FQH state~\cite{Bernevig2012}. Indeed, in the case of a CDW ground state at filling $1/2$, the total number of states in the PES is given by $2 {N \choose N_A}$, which in the previous example gives only $12$ states.

\begin{figure}[htb]
	\includegraphics[width=0.98\linewidth]{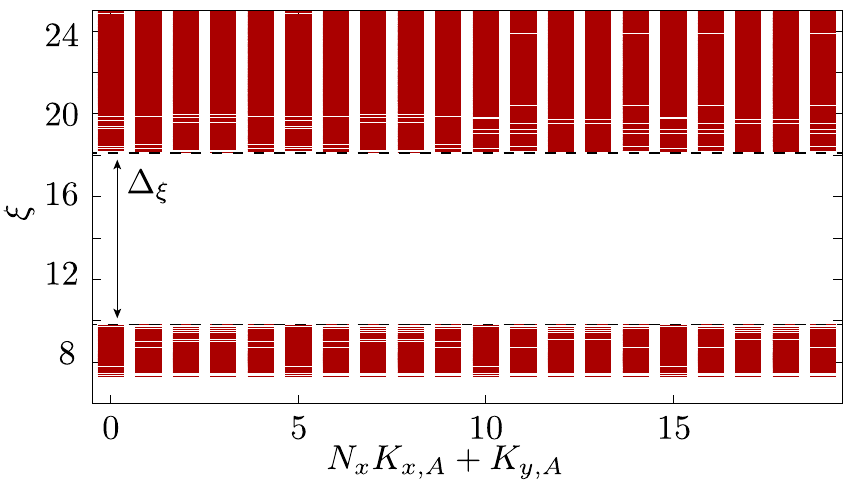}
	\caption{(Color online) Particle entanglement spectrum for $N_B=10$, $N_x=2$, $N_y=10$, $J=0.05E_R$, $l_B=4 \lambda$ and $N_A=5$ as a function of the linearized momentum of a particle in part $A$. The PES shows an entanglement gap $\Delta_\xi$ indicated by the vertical arrow. The number of states below this line is equal to $4004$ which is equal to the total number of quasiholes states with $5$ bosons and $20$ flux quanta for the $\nu=1/2$ Laughlin state. $\Delta_\xi$ marks the entanglement gap as discussed in the main text.}
	\label{fig:ent_spec}
\end{figure}

\begin{figure}[htb]
	\includegraphics[width=0.98\linewidth]{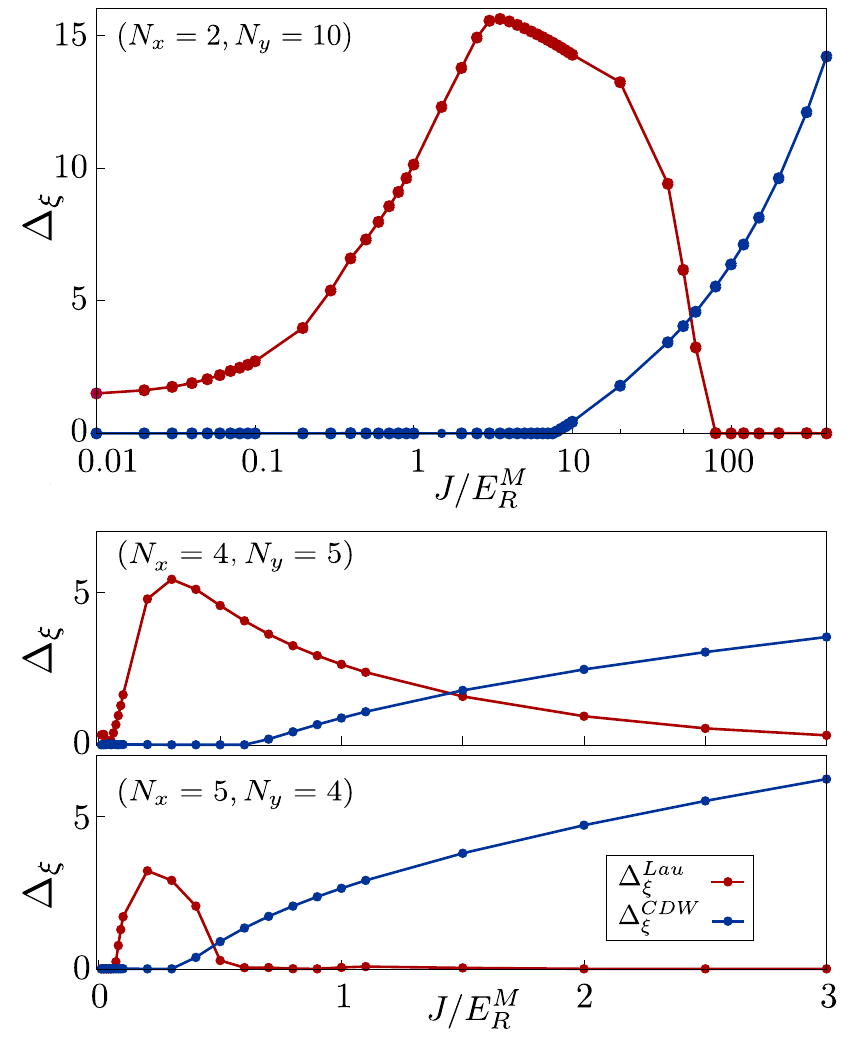}
	\caption{(Color online) Entanglement gap, corresponding to the Laughlin state and CDW state counting, for $\phi=2\pi/a$, $N_B = 10$, $N_A=4$ and $(N_x = 2,N_y = 10)$ (top panel), $(N_x = 4,N_y = 5)$ (middle panel) and $(N_x = 5,N_y = 4)$ (bottom panel) as a function of the inter-wire hopping. Note the logarithmic scale for the horizontal axis in the top panel. In all cases, the Laughlin state emerges for intermediate values of $J$ while at large $J$ a CDW state is found. The $J$ value where this transition occurs, depends strongly on the system size, more precisely on $N_y$, as explained in the main text.}
	\label{fig:entanglement_Gap}
\end{figure}

The entanglement spectrum for the quasi two-fold degenerate ground states of Fig.~\ref{fig:energy_spec} for $N_A=5$ is shown in Fig.~\ref{fig:ent_spec}. It exhibits a clear entanglement gap below which there are $4004$ entanglement levels, which is precisely the expected counting for the $\nu=1/2$ Laughlin state with $5$ particles and $10$ quasiholes. This is a strong evidence for the realization of the Laughlin phase in our model at these parameters. As we did for the energy gap, we can assess the robustness of the FQH phases by looking how the entanglement gap, both the one corresponding to the Laughlin state and the one of the CDW state, changes as the aspect ratio and the hopping strength is changed. This is shown in Fig.~\ref{fig:entanglement_Gap} for $N_B=10$ and $N_A=4$. In agreement with Fig. \ref{fig:energy_spec} and our aforementioned arguments, for $(N_x = 2,N_y = 10)$, the Laughlin gap dominates up to comparably large values of $J$, while for $(N_x = 4,N_y = 5)$ and $(N_x = 5,N_y = 4)$, in decreasing order of $N_y$, it breaks down at smaller and smaller $J$. We note that the maximum entanglement gap of the Laughlin state occurs at increasing $J$ and assumes a larger value as $N_y$ grows. This is in line with our analytical argument that the ground state should converge towards the ideal Laughlin state with infinite entanglement gap for $N_x,N_y, J \rightarrow \infty$.  

\section{Implementation with ultracold atomic gases}

\label{sec:Implementation}

In this Section we discuss the implementation of the semi-continuous Harper-Hofstadter model described
by Eqs.\ (\ref{eqn:ham1}) and (\ref{eqn:hamint}) with ultracold atomic
gases in optical lattices (OL) \cite{Bloch2008,Lewenstein2012}. We
start in Sec.~\ref{sec:planar} with a discussion of the planar
geometry, where the discrete direction is realized as an ordinary
$a=\lambda/2$ lattice with $\lambda$ the optical wavelength of the laser, and coupled by laser assisted tunneling as a Raman process, which allows us to write the position dependent hopping $J_x = J e^{i\phi x}$ with laser phases. In Sec.~\ref{sec:Implementation:wires},
we show how the relevant energy scales, in particular $J$, can be significantly increased
for a double-wire (DW) with subwavelength separation. The construction
of more complex, non-planar trapping geometries like arrays of rings
forming cylinders or tori is discussed in Sec.~\ref{sec:Implementation:Cylindric}.
We show explicitly the implementation of the laser assisted tunneling
in such geometries, providing the possibility to induce a large, tunable
magnetic flux.

\subsection{Planar geometry with optical $\lambda/2$-lattice}

\label{sec:Implementation:OL}

\label{sec:planar} A planar array of parallel wires is readily constructed
with OL potentials (for the implementation of an array of parallel planes see, for example \cite{Gideminas}). Such optical potentials for atomic motion arise as AC Stark shifts in an off-resonant laser beam in a standing wave configuration. In particular, wires with distance $\lambda/2$ are formed by an OL potential of the form $V_{L}(y,z)=V_{y}\sin^{2}(ky)+V_{z}\sin^{2}(kz)$
($k=2\pi/\lambda$). We assume the motion in $z$-direction to be
frozen to the lowest mode by very tight trapping. Hence, the atoms
move freely in the $x$-direction and tunnel with an amplitude $J_{\text{bare}}>0$
between the wires in the transverse $y$-direction {[}see Fig.\:\ref{fig:setup}(a){]}. We note that the assumption of a tight binding model implies $J_{\text{bare}}\ll E_{c}(a)={h^{2}}/{2ma^{2}}$ for a lattice spacing $a$. For an ordinary OL, this energy scale is the {\em recoil energy} $E_{R}=E_{c}(\lambda/2)$ associated with the wavelength of the laser
\cite{Zoller1998}.

A complex interwire hopping $J_{x}=J\exp(i\phi x)$ with a position-dependent
phase is induced with the help of Raman-assisted tunneling \cite{Jaksch2003,Arimondo2010,Aidelsburger2013,Ketterle2013}:
First, the natural tunneling is rendered off-resonant by means of
an energy off-set $\Delta$ between adjacent wires, i.e.~a tilting
of the lattice, created for example by an external magnetic field
gradient. Second, resonant tunneling is restored employing a far-detuned
two-photon Raman process (see Appendix \ref{app:MicroscopicDarkStateHamiltonian}).
As a result, we find the complex Raman assisted hopping $J_{x}$ will be constrained by $|J_{x}|=J<J_{\text{bare}}$ (see also Refs.\:\cite{Arimondo2010,Ketterle2013,Goldman2014}). Its phase $\phi$ is determined by the momentum transfer in $x$-direction during the two-photon Raman process and can be adjusted by variation of the  incidence angle of the two Raman lasers with respect to each other (see Sec.~\ref{sec:Implementation:wires:tunneling} and Appendix \ref{app:MicroscopicDarkStateHamiltonian} for a more detailed description of the laser configurations).  In this Raman scheme the magnetic length $l_B=2\pi/\phi \gtrsim \tilde \lambda/2 $ is  bounded from below by the wavelength $\tilde \lambda$ of the Raman lasers \cite{note4}.

As discussed in Sec.~\ref{sec:Models}, to create quasi-flat
Chern bands we require an inter-wire tunneling amplitude
$J$ on the order of the {\em magnetic (recoil) energy} $E_{R}^{M}=\hbar^{2}\phi^{2}/2m$,
i.e.~$
J/E_{R}^{M}=\mathcal{O}(1)
$.
In practice, this ratio should be tunable in a range from
$J/E_{R}^{M}\lesssim0.5$ to $J/E_{R}^{M}\gtrsim2$. It is not desirable
to increase this ratio by simply decreasing $E_{R}^{M}$, as this
would increase the magnetic length $l_{B}=2\pi/\phi$ and thus reduce
the particle density at a given filling along with the relevant many-body
interaction energies. Hence, the larger $J$ the better, which is, however,
bounded from above by $J_{\text{bare}}\ll E_{R}$. In summary, using
ordinary OLs, we are constrained by the set of inequalities 
\begin{align}
E_{R}^{M}\sim J<J_{\text{bare}}\ll E_{c}(a=\lambda/2)=E_{R}\;,
\end{align}
and as a realistic compromise, a possible choice of parameters is
$J=0.05E_{R},l_{B}=4\lambda$. Then, the ratio $J/E_{R}^{M}=0.8$
and the flatness ratio of the lowest band is $137$.

From an experimental viewpoint, larger energy scales $E_{R}^{M}\sim J\sim E_{R}\gg k_{B}T$
can be crucial due to finite temperature $T$ (see also Sec.~\ref{sec:Models:twoWires}). It is thus desirable to design schemes with subwavelength interwire spacing. We will discuss this in the following Subsection for the case of a DW. 

\subsection{Enhanced energy scales in a subwavelength double wire}

\label{sec:Implementation:wires}

In the following, we demonstrate the enhancement of the
Raman assisted interwire hopping (see also Sec.~\ref{sec:Implementation:wires:tunneling})
in a DW setup with wire separation $a=\ell\ll\lambda/2$
far below the laser wavelength, where 
\begin{align}
a=\ell\ll{\lambda}/{2}\;\Leftrightarrow\;E_{c}(a=\ell)\gg E_{R}\;.
\end{align}

The construction (Sec.~\ref{sec:Implementation:wires:construction}) is based on the nanoscale `dark state' optical potentials
 introduced in Ref.\:\cite{Zoller2016} (see also Ref.\:\cite{Gorshkov2016}).
We will show that Raman assisted hopping amplitudes for typical parameters can reach values $J\sim E_{R}$ (Sec.~
\ref{sec:Implementation:wires:tunneling}),
more than one order of magnitude larger than those obtained in standard OL setups \cite{Ketterle2013,Aidelsburger2013}. Furthermore, this
represents by no means a fundamental upper limit for setups featuring
subwavelength structures. In parallel to this enhancement of the Raman
assisted interwire hopping, the magnetic flux has to be increased
to satisfy the requirement $J/E_{R}^{M}=\mathcal{O}(1)$. In our concrete example,
a magnetic length $l_{B}\sim\lambda$ is sufficient to obtain $E_R^M \sim J\sim E_{R}$
(see below). This can be achieved by ordinary Raman assisted tunneling
techniques (see Sec.~\ref{sec:Implementation:wires:tunneling}).

\subsubsection{Subwavelength double wire}

\label{sec:Implementation:wires:construction}

\begin{figure}[t]
\includegraphics[width=86mm]{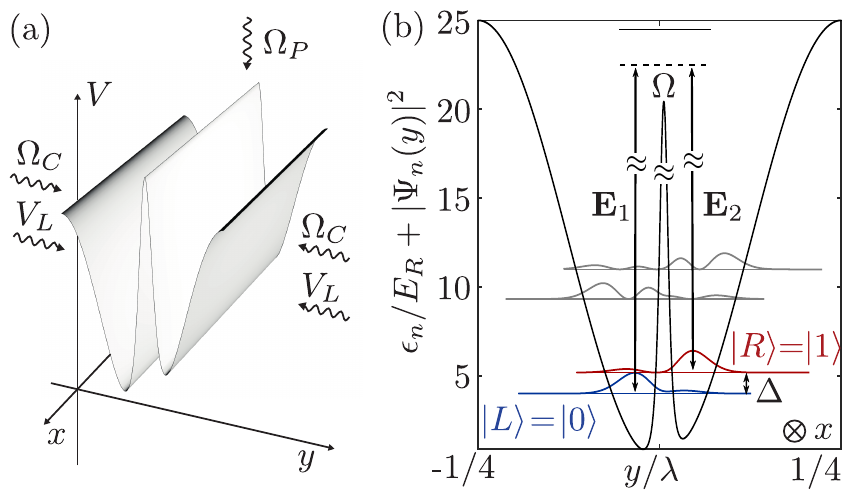} \caption{(Color online) Two parallel wires with subwavelength separation $\ell\ll\lambda/2$
are created by inserting a nanoscale optical potential barrier into
a potential well $V_{L}(y)$ generated by a familiar off-resonant
OL. Panel $(a)$ sketches the potential landscape with a pair of laser
counterpropagating beams creating $V_{L}$. Two further lasers with
Rabi frequencies $\Omega_{P}$ and $\Omega_{C}$ create the nanoscale
barrier. Panel $(b)$ shows the potential and wavefunctions at fixed
position in $x$-direction. Note the small displacement $y_{0}\sim0.1a_{H}$ of the
barrier, w.l.o.g.\ towards positive $y$, that leads to the formation
of distinct left- and right states (resp.\ ground $\ket{0}=\ket{L}$
and first excited state $\ket{1}=\ket{R}$) separated by an energy
offset $\Delta$. Via far detuned Raman transitions with frequency
detuning $\delta\omega\approx\Delta/\hbar$, resonant tunneling is
restored and a spatially varying phase is induced due to a finite
momentum transfer $\delta k_{x}=k_{1,x}-k_{2,x}$ in $x$-direction. }
\label{fig: nanoscale _wires} 
\end{figure}

A DW separated by a subwavelength distance is obtained using the nanoscale `dark state' optical potentials introduced recently
in Ref.\:\cite{Zoller2016} (see also Ref.\:\cite{Gorshkov2016}). Inserting such a barrier into a potential
well, created, for example, as a 1D optical wire in a standard OL setup, we can split this wire into two subwires (see Fig.\:\ref{fig: nanoscale _wires}(a)). These
subwires feature a spatial separation far below the laser wavelength,
comparable to the typical size of the groundstate wave function of
the potential well of the original OL  (see Fig.\:\ref{fig: nanoscale _wires}(b)).

To produce a DW configuration extending in the $x$-direction, we
consider, in the $y$-direction, a single potential well of a standard
1D off-resonant OL $V_{L}(y)=V_{y}\sin^{2}\left(ky\right)$ with height
$V_{y}$ (see Fig.\:\ref{fig: nanoscale _wires}). In the harmonic
approximation, this potential well is characterized by the trapping
frequency $\hbar \omega_{T}\approx\sqrt{4V_{y}E_{R}}$ and by the size of
the ground state wave function $a_{H}=\sqrt{\hbar/(2m\omega_{T})}$.
Following reference \cite{Zoller2016}, the nanoscale barrier is added
as a nonadiabatic optical potential 
\begin{align}
V_{na}(y)=E_{R}\frac{\epsilon^{2}\cos^{2}(k(y-y_{0}))}{\left[\epsilon^{2}+\sin^{2}(k(y-y_{0}))\right]^{2}}
\end{align}
for the dark state in an atomic $\Lambda$-system created using two
additionally applied lasers, a moving wave with Rabi frequency $\Omega_{P}$
constant in the $x$-$y$-plane and a standing wave $\Omega_{C}(y)=\Omega_{C}\sin(k(y-y_{0}))$
(see the Appendix\:\ref{app:MicroscopicDarkStateHamiltonian} for
details on the atomic level scheme). Hereby, $\epsilon=\Omega_{P}/\Omega_{C}\ll1$
is an external parameter, controlled by the lasers. It determines
the subwavelength width ($\ell\sim\epsilon/k\ll\lambda/2$) and the
height ($\sim E_{R}/\epsilon^{2}$) of the barrier which appears around
$y_{0}$. Physically, this potential barrier is caused by the rapid
change of the identity of the dark state in the subwavelength region
of width $\ell$, see the Appendix \ref{app:MicroscopicDarkStateHamiltonian}
and Ref.\:\cite{Zoller2016} for details. Ignoring the frozen motion
in the $z$-direction for which we assume tight confinement, the dark
state Hamiltonian takes the form 
\begin{align}
H_{D}(x,y)=-\left(\frac{\hbar^{2}\partial_{x}^{2}}{2m}+\frac{\hbar^{2}\partial_{y}^{2}}{2m}\right)+V_{L}(y)+V_{na}(y)\:.\label{eq: HDarkstate}
\end{align}
For $\epsilon\ll1$, the eigenstates of $H_{D}$ are grouped into
pairs of states with small energy offset $\Delta\ll\hbar\omega_{T}$.
Different pairs are separated by a large gap $E_{\text{gap}}\approx2\hbar\omega_{T}$.
In the following, we can hence restrict to the lowest pair formed
by ground $\ket{0}$ and first excited state $\ket{1}$. Due to a
small displacement $y_{0}\ll a_{H}$ of the barrier with respect to
the center of the harmonic trap, one of these two states (as in every
pair) is mostly localized on the left, one on the right of the barrier,
with only a small fraction leaking through. We hence identify a DW
system consisting of distinct left $\ket{L}\equiv\ket{0}$ and right
$\ket{R}\equiv\ket{1}$ wire, respectively (for the configuration
shown in Fig.\:\ref{fig: nanoscale _wires} with the barrier displaced
to the right, i.e.\ $y_{0}>0$). We are interested in a situation
where the wires are well separated. This means that the localization
parameter $\text{Loc}\equiv1/2\left(P_{L}+P_{R}\right)$ with $P_{L}\equiv\int_{-\infty}^{y_{0}}\text{d}y|\braket{y|L}|^{2}$,
$P_{R}\equiv\int_{y_{0}}^{\infty}\text{d}y|\braket{y|R}|^{2}$ has
to be close to unity. In the following, we hence keep $\text{Loc}>0.9$,
balancing between the spatial separation of the wires and the strength
of the interwire coupling (see below) related to the spatial overlap
of their wave functions.

Summarizing, we have created a double wire with interwire distance
$\ell\ll\lambda/2$ far below the laser wavelength. This provides
the opportunity to obtain Raman-assisted interwire couplings much
larger than in a familiar optical lattice, because the {\em  a priori} bare hopping amplitudes can be much larger than for a standard OL,  as shown in the following
subsection.

\subsubsection{Enhanced Raman-assisted tunneling amplitude}

\label{sec:Implementation:wires:tunneling}

We now discuss a Raman-assisted tunneling scheme for the subwavelength
DW, where complex tunneling phases can be obtained in a Raman assisted tunneling scheme as described above. Our goal is to maximize the Raman assisted
tunneling amplitude $J$ and to achieve simultaneously a large magnetic
flux $E_{R}^{M}/J=\mathcal{O}(1)$. We compare to previous results
obtained in ordinary OLs \cite{Aidelsburger2013,Ketterle2013}.

The tunnel coupling between the wires is generated by two additional
Raman beams $\mathbf{E}_{1}$, $\mathbf{E}_{2}$ with frequencies
$\omega_{1}$, $\omega_{2}$ and wave vectors $\mathbf{k}_{1}$, $\mathbf{k}_{2}$
located in the $x$-$y$-plane. They drive a far-detuned two-photon
transition via an auxiliary level (Figure \ref{fig: nanoscale _wires},
detailed in Appendix \ref{app:MicroscopicDarkStateHamiltonian}).
Adiabatic elimination of the auxiliary level leads to an additional
time-dependent driving term 
\begin{align}
H_{RM}(x,y,t)=\Omega\cos\left(\delta k_{x}x+\delta k_{y}y-\delta\omega t\right),\label{eq: HRaman}
\end{align}
in the full Hamiltonian $H(t)=H_{D}+H_{RM}(t)$, where $\Omega$ is
the effective two-photon Rabi frequency, $\delta\omega=\omega_{1}-\omega_{2}$
the frequency detuning and $\delta\mathbf{k}=\mathbf{k}_{1}-\mathbf{k}_{2}$
the momentum transfer. Choosing $\delta\omega\approx\Delta/\hbar$,
$H_{RM}$ induces resonant tunneling between the two wires whose amplitude
is in the perturbative regime $\Omega\ll\hbar  \delta\omega$ given by 
\begin{align}
J_{x}\approx\frac{\Omega}{2}\braket{L|e^{-i\delta k_{y}\hat{y}}|R}e^{-i\delta k_{x}x}\equiv Je^{i\phi x}.\label{eq: Jpert}
\end{align}
Importantly, the matrix element $|\braket{L|e^{-i\delta k_{y}\hat{y}}|R}|=\mathcal{O}(1)$
is increased in our setup by at least one order of magnitude compared
standard OLs where we have $|\braket{L|\exp({-i\delta k_{y}\hat{y}}|R})|\sim J_{\text{bare}}^{\text{OL}}/E_{R}\lesssim\mathcal{O}(1/10)$
between any two adjacent wires $L$ and $R$ \cite{Ketterle2013}.
This is caused by the larger spatial overlap of the wave functions
and leads to the strongly enhanced tunneling amplitudes in the subwavelength
DW.

\begin{figure}[t]
\includegraphics[width=86mm]{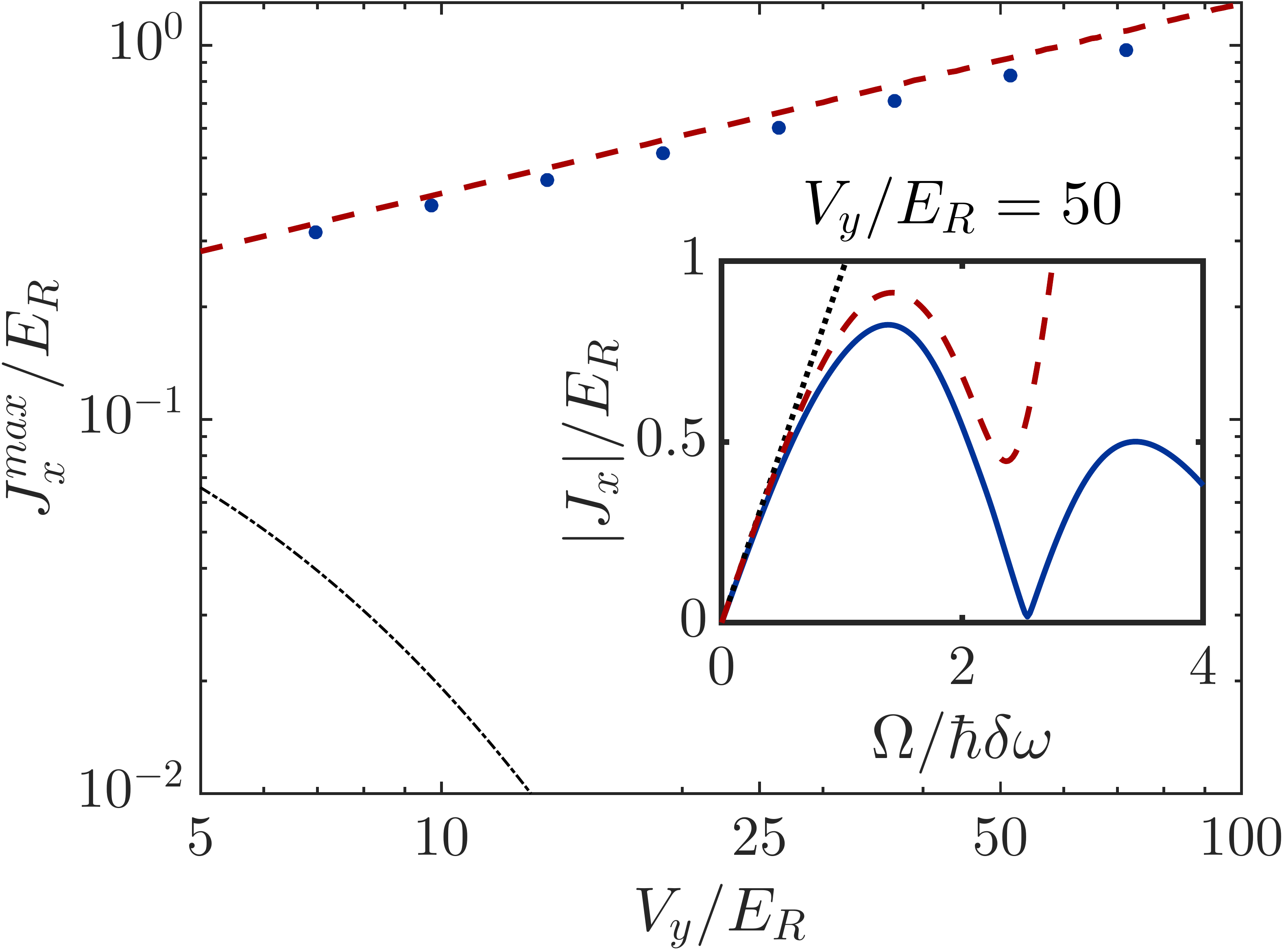}
\caption{(Color online) Maximal achievable Raman induced tunneling rate between two atomic nanowires
with subwavelength separation as a function of the strength of the
external trapping potential $V_{y}$ (Blue dots: numerical Floquet
analysis, Red dashed line: second order perturbation theory in an
expansion of an effective Hamiltonian in powers of $\Omega/\hbar \delta\omega$).
For comparison, the bare tunneling rate of a standard OL is shown (black dashed-dotted line),
bounding the Raman assisted hopping in the corresponding tilted OL
\cite{Ketterle2013,Goldman2015}. Inset: tunneling rate as a function
of the driving amplitude, the two-photon Rabi frequency $\Omega$
(Numerical Floquet analysis (blue), zeroth (black dotted) and second
(red dashed) order perturbation theory). $\epsilon\approx0.1$, $y_{0}\approx0.1a_{H}$,
$\delta k_{y}=2k$ and $\delta\omega=1.1\Delta/\hbar$ have been chosen
to maximize resonant tunneling. }
\label{fig: coupling} 
\end{figure}

Our goal is now to find the maximum possible Raman assisted tunneling
amplitude (c.f.~Fig.\:\ref{fig: coupling}).
Here, we investigate the overall behavior of the tunneling amplitude
as a function of the driving strength $\Omega$. This requires to
access the non-perturbative strong driving regime using a numerical
Floquet analysis. As shown in the inset of Fig.\:\ref{fig: coupling},
the tunneling amplitude follows the common trend \cite{Arimondo2010,Ketterle2013,Goldman2014}:
A linear growth with the driving in the perturbative regime before
the onset of higher order effects diminishing the coupling for strong
driving $\Omega\gtrsim\hbar \delta\omega$. It is renormalized due to time-dependent
AC-Stark shifts of the individual levels $\ket{L}$ and $\ket{R}$.
For details on this higher order effects, a systematic perturbation
theory in powers of $\Omega/\hbar\delta\omega$ and the numerical procedure,
we refer to Appendix \ref{app:FloquetTwoCoupledWires}. 

Due to the large tunneling matrix element in Eq.~\eqref{eq: Jpert},
the maximum amplitude of the tunnel coupling between the wires is
strongly enhanced compared to the results of Ref.\:\cite{Ketterle2013}
for the standard setup with a tilted OL. As an example, for the tilted
OL with the amplitude $V_{0}=9E_{R}$, the amplitude of the Raman
assisted tunneling between the nearest sites is bounded by the bare
tunneling $J_{b}=0.02E_{R}$ \cite{Ketterle2013}. Using the same
OL with the subwavelength barrier placed in one of its wells, one
can achieve an order of magnitude increase, $J_{x}\approx0.2E_{R}$,
of the tunneling amplitude (see Fig.\:\ref{fig: coupling}).

In contrast to standard OLs, the Raman assisted tunneling amplitude
in our setup can be further enhanced by increasing the amplitude of
the trapping potentials $V_{y}$, as shown in Fig.\:\ref{fig: coupling}.
This results in the increase of the level separation $\Delta$ leading
to a larger value $\Omega_{\textrm{max}}\sim\hbar \delta\omega\sim\Delta$
where the maximum tunneling amplitude $J_{x}\sim\Omega_{\textrm{max}}$
is reached. This overcomes a decrease of the tunneling matrix element
$\braket{L|e^{-i\delta k_{y}\hat{y}}|R}$. Quite remarkably, the tunneling
amplitude $J_{x}$ reaches the values of $\sim\! 1 E_{R}$ for $V_{y}=100E_{R}$
(c.f.~Fig.\:\ref{fig: coupling}).

At the same time, a large flux $(\hbar^{2}\phi^{2}/2m)/J=\mathcal{O}(1)$
is achieved using the following laser configurations: We take laser
$\mathbf{E}_{1}$ to be $\sigma_{+}=-({\mathbf{e}_{z}}+i{\mathbf{e}_{x}})/\sqrt{2}$
polarized and propagating along the $y$-direction. The laser $\mathbf{E}_{2}$
propagates the $x$-$y$-plane with an incidence angle $\theta$ with
respect to $\mathbf{E}_{1}$. The momentum transfer during the two-photon
process is thus $\delta\mathbf{k}=\mathbf{k}_{1}-\mathbf{k}_{2}$.
By variation of $\theta$, we can hence adjust the phase $\phi=\delta k_{x}$,
up to large fluxes $(\hbar^{2}\phi^{2}/2m)/E_{R}=\mathcal{O}(1)$
(for $\theta\approx\pi/2$).

To summarize, our discussion shows
that tunneling amplitudes $J\sim E_{R}$ can be reached in a subwavelength DW,
which are enhanced by more than one order of magnitude compared to familiar
OLs. In addition, this scheme allows one to achieve fluxes $E_{R}^{M}\sim J$ compatible with the enhanced hopping.

\subsection{Non-planar trapping geometries: cylinder and torus}

\label{sec:Implementation:Cylindric}

Below we describe the realization of optical potentials for ultracold
atoms with cylindric rotational symmetry — the cylinder consisting
of an array of rings (Sec.~\ref{sec:Implementation:Cylindric:cylinder})
and the torus (Sec.~\ref{sec:Implementation:Cylindric:torus}).
The construction is based on Laguerre-Gaussian (LG) laser beams.
Similar beams allow also to thread artificial magnetic field flux
through the surface of each geometry and realize the scenario of the
flat band Hamiltonian\:\eqref{eqn:ham1}. All beams are set deep in the
paraxial regime, where beam waist $w_{0}$ satisfy $w_{0}\gg\lambda$,
and the constructed lattices will feature standard interwire separations
of $a=\lambda/2$.

\subsubsection{Cylindric array of rings}

\label{sec:Implementation:Cylindric:cylinder} The cylinder geometry
{[}see Fig.\:\ref{fig:setup}(b){]} can be naturally obtained using
a laser mode with a cylindric symmetry, such as a LG beam (see Refs.\ \cite{Allen1992,Allen2003,Padget2008,Zeilinger2012,Andrews2011,Andrews2012}
for theory description and Refs.\:\cite{Thirugnanasambandam2010,Senatsky2012,Zupancic2013,Preiss2015}
for the experimental state of the art). The spatial mode of the LG
beams is indexed by integers $l$ and $p$ (we consider here $p=0$)
which we refer to as $LG_{p,l}$. Its electric field amplitude is
\begin{equation}
\mathbf{E}_{0}(\rho,\varphi,y)={\pmb{\bm{\epsilon}}}_{0}\,Ef_{p,l}(\rho,y)e^{il\varphi}e^{iky},\label{eq:LGparaxial}
\end{equation}
where the $\mathbf{{\epsilon}}_{0}$ is the polarization vector, taken
to be $\sigma_{+}=-({\mathbf{e}_{z}}+i{\mathbf{e}_{x}})/\sqrt{2}$.
Here $E$ denotes the field strength. We have chosen cylindrical coordinates
with azimuthal angle $\varphi$ and $\rho=\sqrt{x^{2}+z^{2}}$ the
distance to the symmetry axis ${\mathbf{e}_{y}}$. The rotationally
invariant field amplitude is 
\begin{equation}
f_{p,l}(\rho,y)=\frac{w_{0}}{w_{y}}\sqrt{\frac{2p!}{\pi(p+|l|)!}}\xi^{|l|}L_{p}^{|l|}(\xi^{2})e^{-\xi^{2}/2},\label{eq:LGfpl}
\end{equation}
with $\xi=\sqrt{2}{\rho}/{w_{y}},$ $w_{y}=w_{0}\sqrt{1+\left({y}/{y_{R}}\right)^{2}},$
and $w_{0}$ is the paraxial waist of the beam. Further, the Rayleigh
range $y_{R}={w_{0}^{2}\pi}/{\lambda}$ and $L_{p}^{|l|}$ is a generalized
Laguerre polynomial (for $p=0:L_{0}^{l}(\xi)=1)$. The plane $y=0$
in this description is the focal plane of the LG beam.

\begin{figure}[t]
\includegraphics[width=86mm]{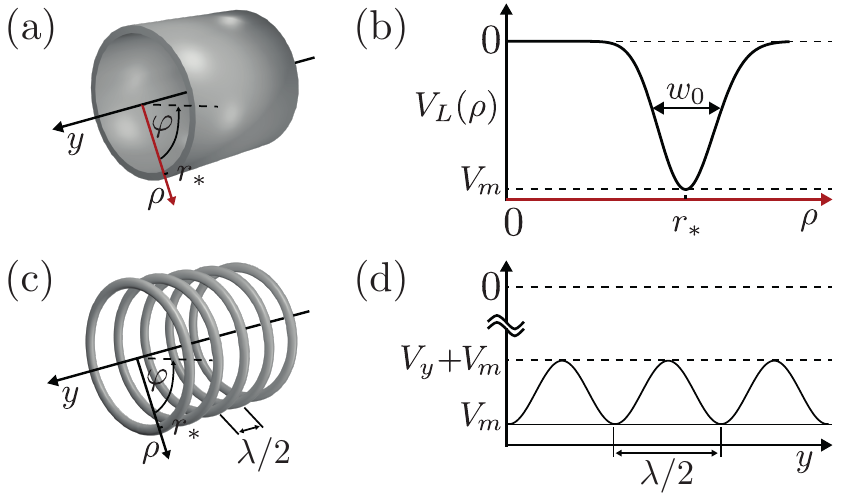} \caption{(Color online). Panel (a) shows the rotationally invariant, cylindrical
potential well created by a Laguerre-Gaussian beam $LG_{0}^{l}$  in cylindrical coordinates.
Panel (b) shows the intensity profile in the radial direction, featuring
a Gaussian potential centered at finite radius $r_{*}=w_{0}\sqrt{l/2}$
 of width $w_{0}$ and depth $V_{m}$. Panel (c) shows the slicing
of the cylindric potential featured in (a) by an extra standing wave
potential creating a series of rotationally invariant parallel rings.
Panel (d) shows the $y$-dependence of the resulting potential. }
\label{fig:figLG} 
\end{figure}

The LG beam creates a potential {[}see Fig.\:\ref{fig:figLG}(a){]}
\begin{equation}
V_{L}(\mathbf{r})=V_{m}\;|f_{p,l}(\rho,y)|^{2},\quad V_{m}<0.\label{eqn:Vring}
\end{equation}
which in radial direction contains a Gaussian minimum located at finite
position of $r_{*}=w_{0}\sqrt{{l}/{2}},$ thus realizing a potential
well corresponding to the side surface of a cylinder, when $y/y_{R}\ll1$
{[}see Fig.\:\ref{fig:figLG}(b){]}. 

On top of the cylindrical potential we add a lattice potential $V_{L}(y)=V_{y}\sin^{2}(ky)$
splitting the cylindric landscape into a series of parallel ring traps
separated by $a=\lambda/2$ distance {[}see Fig.\:\ref{fig:figLG}(c),(d){]}.
It should be noted that the LG beams may be used in a fashion similar
to Sec.~\ref{sec:Implementation:wires:construction} to create
a subwavelength barrier with ring shape and in the end the two parallel
rings with subwavelength separation $\ell\ll\lambda$.

\subsubsection{Raman-assisted tunneling}

\label{sec:Implementation:Cylindric:raman} Below we discuss specific
features of the Raman-induced hopping in the cylindric geometry. We
start by explaining the creation of a synthetic magnetic field. Then,
we discuss how to couple many rings using just two lasers, addressing
residual focusing effects. They define a maximal achievable total
size of the system, allowing to couple together several tens of rings
for which the Hamitonian\:\eqref{eqn:ham1} is realized.

The Raman scheme for laser-assisted hopping proceeds in analogy to
the one discussed in Section~\ref{sec:Implementation:wires}. First, a magnetic
field gradient creates a local energy offset between neighboring rings.
Second, the rings are coupled via a Raman process tuned to that offset.
The non-standard part is however the choice of the LG laser beams
for the lasers creating the two-photon transition.

We employ two $\sigma_{+}$ polarized LG beams: $LG_{0,l_{1}}$ and
$LG_{0,l_{2}}$, counterpropagating in the $y$-direction. The azimuthal
complex phase carried by the LG beams {[}see Eq.\:\eqref{eq:LGparaxial}{]}
amounts to a total angular momentum transfer during the two-photon
process. In analogy to Eq.\:\eqref{eq: Jpert}, this leads to a tunneling
phase of $\phi=(l_{1}-l_{2})/R$ where $R$ denotes the radius of
the cylinder. The fundamental limit for the complex phase $\phi$
is the diffraction limit, which is bounded by $\phi^{-1}>\phi_{c}^{-1}=\lambda_{R}/2$
with $\lambda_{R}$ the wavelength of the Raman lasers \cite{Zoller2016a}.
Even far from the diffraction limit, for $\phi/\phi_{c}=1/8$, we
obtain $\hbar^{2}\phi^{2}/2m=1/16E_{R}$ and hence $J/(\hbar^{2}\phi^{2}/2m)=\mathcal{O}(1)$
in the desired regime (for realistic $J=0.05E_{R}$, see Section\:\ref{sec:Models}).

To simultaneously Raman-couple many wires resonantly using just two lasers with
the desired amplitude $J$, one needs to assure that a global choice
of detuning of the two lasers $\delta\omega=\omega_{1}-\omega_{2}$
{[}as discussed in Section\:\ref{sec:Implementation:wires:tunneling}{]}
can be made. In the cylindric geometry case, the local
energies $E_y$ of the rings are determined by the cylindric optical potential \eqref{eqn:Vring}, by the trap frequency $\hbar \omega_T\approx \sqrt{4V_yE_R}$ of the lattice potential $V_L(y)$ and by an additional external magnetic field gradient giving rise to an offset $\Delta$ between adjacent rings. They read as
\begin{align}
E_{y}=V_{m}\left[1-\frac{y^{2}}{y_{R}^{2}}+\frac{y^{4}}{y_{R}^{4}}+{\mathcal{O}}\left(\frac{y^{6}}{y_{R}^{6}}\right)\right] + \Delta \frac{y}{a} + \hbar \omega_T \;. \label{eqn:focusingOffsets} 
\end{align}
Hence, the total difference between neighbouring rings is $\Delta_{y}=E_{y+a}-E_{y}=\Delta+\delta_{y}$ where  $\delta_{y}$
depends non-linearly on $y/y_R$. This is in contrast to the planar
case where $\Delta_y=\Delta=\text{const.}$, i.e.\ $\delta_y=0$. 

In the effective Hamiltonian\:\eqref{eqn:ham1}
the presence of $\delta_{y}\neq0$ adds an extra local energy term
$\sum_{y}\tilde \delta_y \hat{n}_{y}$ where $\tilde{\delta}_y=\sum_{y'\geq y} \delta_y$. 
To keep this extra term negligible, we hence require that 
\begin{align}
\tilde \delta_y \ll J\label{eqn:ringTiltCondition} \; .
\end{align}
This defines the maximal number
$\sim y_{R}\sqrt{J/V_{m}}/a$ of lattice rings to be resonantly coupled. 
This
number is enlarged if the quadratic part $\sim y^2/y^2_R$ in equation  \eqref{eqn:focusingOffsets}
is compensated by an additional anti-trapping potential, allowing
$\sim y_{R}\sqrt[4]{J/V_{m}}/a$ lattice rings to be coupled simultaneously
by a Raman process. Here $y_{R}$ is limited by the separation between radial, transverse modes of $\Delta E_{\rho}=\sqrt{8\hbar^{2}V_{m}/mw_{0}^{2}}$ where $w_{R}$ is the waist of the LG beams creating the cylinder.  For realistic values of parameters one can construct
20 or more parallel rings.

\begin{figure}[t]
	\includegraphics[width=86mm]{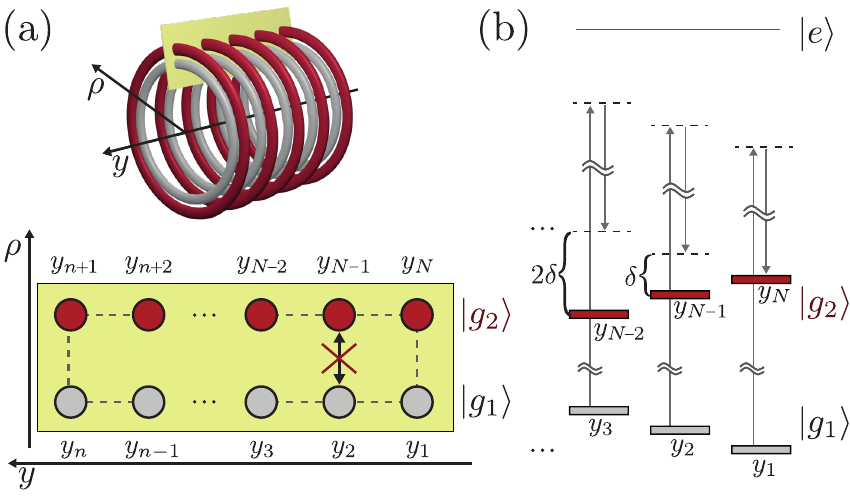} \caption{(Color online). Torus geometry created by Raman-coupling ends of two
		coaxial cylinders. Panel (a) shows coupled rings that together create
		a toric geometry. Red and silver color indicate lattice sites occupied
		by atoms in atomic states $g_{1}$ and $g_{2}$ respectively. Panel
		(b) shows the landscape of local energies of sites in the two rings which is created by a
		 single external magnetic field gradient. It features different tilting for the individual internal states (red as opposed to silver). This allows to resonantly
		Raman-couple just two rings [$y_{1}$ and $y_{N}$ in panel (b)].}
	\label{fig:figLG2} 
\end{figure}

\subsubsection{Torus geometry}

\label{sec:Implementation:Cylindric:torus} To realize a torus geometry of coupled rings,
we create two copies of cylinders nested into each other.
Only at the rings located at the edges, inner and outer cylinder are coupled (by laser-assisted tunneling), closing the geometry to a torus  {[}see Fig.\:\ref{fig:setup} and \ref{fig:figLG2}{]}.
To suppress
the hopping between the cylinders at other positions,
the cylinders are implemented as state-dependent potentials for ultracold
atom gas populating two hyperfine ground states $g_{1}$ and $g_{2}$  \cite{Gerbier2010,Yi2008,Belmechri2013}.
The potential felt by each atomic state contains just a single cylinder.
Any hopping of atoms between the two cylinders involves a
change of atomic configuration which prohibits the occurrence of (bare)
tunneling processes.
By a specially constructed laser-assisted
hopping Raman scheme \cite{Aidelsburger2013,Ketterle2013}, inner and outer cylinder can be coupled. Crucially, the state dependent potentials
allows us to keep the spatial separation of the cylinders
$\sim\lambda$, which is necessary for a non-negligible hopping amplitude
between the two rings (see Eq. \eqref{eq: Jpert}).

In more detail, the laser-assisted tunneling scheme coupling only rings at the edge is constructed as follows. Since each cylinder is populated by a different atomic state, the tilts
of the local energies due to external magnetic field gradient 
can be chosen differently {[}see two gradients of energies for silver and red
states in Fig.\:\ref{fig:figLG2}(b){]}. Two rings indexed as $j$ and $N-j+1$ have the same position $y_j=y_{N-j+1}$ in
$y$-direction and are concentric. The difference
of local energies of the outer and inner ring $\Delta_{j}=E_{N-j+1}-E_{j}$
 depends nontrivially on the position $y_j=y_{N-j+1}$. When engineering
a Raman process, it is therefore possible to adjust the relative detuning
of the Raman lasers $\delta\omega$ to be resonant only with a single
$\Delta_{j}$ and off-resonant for all others  {[}see Fig.\:\ref{fig:figLG2}(b){]}. 
This allows to  couple inner and ring only at the edges of the cylinder, i.e.\ at positions $y_{j=1}=y_{j=N}$ and $y_{j=N/2}=y_{j=N/2+1}$, respectively. In this way, the  two cylinders  are closed into a torus geometry. For further details, we refer to Appendix~\ref{app:spinDependentTorus}.

\section{Conclusions}
In this work we have proposed and studied the realization of a {\em semi-continuous Harper-Hofstadter model}  with cold atoms in optical lattices. This model is defined in the semi-continuous position space of an array of wires. It hence naturally combines key physical aspects of recently observed lattice systems, such as the Harper-Hofstadter (HH) model \cite{Jaksch2003,Aidelsburger2013} and the two leg ladder \cite{Atala2014}, with continuum systems such as Landau levels and rotating condensates \cite{Cooper2001}, respectively.

In the continuous lowest Landau level, both the energy dispersion and the Berry curvature are exactly flat, and the energy gap to the next higher Landau level is determined by the magnetic cyclotron frequency $\omega_c=eB/m$, i.e.~proportional to the magnetic field $B$. Here instead, in Hamiltonian~(\ref{eqn:hky}), there are two {\emph{independent} knobs for tuning (i) the magnetic length by varying the synthetic flux $\phi=2\pi/l_B$ , and (ii) tuning the gap along with the flatness by varying the hopping strength $J$. Both these knobs are readily accessible experimentally in present experiments with ultracold atomic gases, thus allowing to realize various parameter regimes. 

This situation is also strikingly different from the HH on a lattice, where the flatness of the lowest (Chern) band is determined by the magnetic flux per plaquette rather than by the hopping strength, and flat bands resembling the lowest Landau level are achieved in the limit of large magnetic length. To overcome this issue, Ref. \cite{KapitMueller} proposed an extension of the HH model with flat bands at the expense of longer range hopping, and subsequently quite some effort has been devoted to identifying various mechanisms that generate approximately flat Chern bands \cite{Bergholtz2013,NathanReview}. In this context, the semi-continuous HH model provides a particularly easy route towards flat Chern bands with ultracold atomic gases, where the flatness is simply tuned by the nearest neighbor hopping $J$.  

As an illustration of how FQH states can be stabilized in the coupled wire setting, we have shown with both analytical arguments and a numerical analysis the emergence of a Laughlin phase in a system for experimentally realistic parameters. In the limit of a small $J$ acting as a perturbative coupling between strongly correlated wires (Luttinger liquids), it is well known from renormalization group arguments how various more complex FQH states can be appear in coupled wires \cite{TeoKane2014}. In the present atomic setting with flat Chern bands in the limit of large $J$, we identify the search for specific longer range interactions stabilizing non-Abelian FQH states in arrays of coupled wires, and their experimental realization with the cold atom toolbox as an interesting subject of future research.

We note that synthetic dimensions, where an internal degree of freedom of the atom \cite{Celi2014} or harmonic oscillator states of a confining potential \cite{Goldman2016} are interpreted as the discrete dimension, may provide an alternative route towards experimentally realizing coupled wires. Regarding quantum Hall physics, a first study in this context \cite{SaitoPreprint} indicates that no FQH states are stabilized there due to the strongly anisotropic nature of the two-body interactions. A possible way around this issue is provided by the approach of using the angular variable of a photonic ring cavity \cite{Ozawa2017} as a synthetic dimension, which results in spatially local interactions.

For the double-wire -- the minimal setting hosting a synthetic magnetic flux -- we have proposed the realization of a pair of atomic nanowires with sub-wavelength separation, which significantly increases the relevant energy scales over existing OL realizations.  At effective temperatures already realized in state of the art cold atom experiments, we argue that this crucial enhancement enables the visibility of intriguing quantum phenomena in Fermi gases such as Lifshitz transitions.  Generalizing these techniques to the case on more than two wires would open up entirely new possibilities for the experimental observation of strongly correlated phenomena.

The use of Laguerre-Gauss beams allows us to create non-planar coupled wire systems in cylinder- disk- and torus-geometries. A counterpart of this construction for the HH lattice model has recently been presented in Ref. \cite{Zoller2016a}. We conclude by stressing another key advantage of the present semi-continuous HH model as compared to its lattice counterpart, in the context of such non-planar geometries. In the lattice version, we proposed to use tightly focused beams (see Ref.\:\cite{Zoller2016a} and references cited there) in order to achieve lattice constants in the circumferential direction on the order of optical wavelengths. This was important to achieve energy scales -- in particular band gaps -- that are on the order of those found in conventional OLs. Here, as such band gaps are simply tuned by the hopping $J$ in the discrete axial direction (see Fig. \ref{fig:setup} (b)), where the lattice spacing is anyway (at most) the one of a conventional OL, there is no need for the additional experimental effort of tight focusing.

\label{sec:Conclusions}

\section*{Acknowledgments}
We thank E. Bergholtz, J. Dalibard, N. Goldman, and H. Pichler for discussions and feedback on the manuscript. Work at Innsbruck is supported by the European Research Council (ERC) Synergy Grant UQUAM, the Austrian Science Fund through SFB FOQUS (FWF Project No.\ F4016-N23), and EU FET Proactive Initiative SIQS. A.S.\ acknowledges funding by the European Research Council (ERC) grant WASCOSYS (No.\ 636201).

\appendix

\section{Band flatness and Berry curvature of the Hamiltonian \eqref{eqn:ham1}}

\label{app:BandFlatness}

In this appendix we provide further details on the flatness ratio
and Berry curvature of the band structure, and the localization properties
of the edge modes associated with the Hamiltonian \eqref{eqn:ham1}.
The characteristic energy scales are the coupling $J$ and the magnetic
recoil energy $E_{R}^{M}=\hbar^{2}\phi^{2}/2m$.

The flatness of the bands (as described in Sec.~\ref{sec:Implementation:Cylindric:cylinder})
is described by the ratio $F\equiv BG/BW$ of the bandwidth $BW$
of the lowest Bloch band to the band gap $BG$ between the lowest
and the first excited band. In the limit $J/E_{R}^{M}\gg1$ we obtain
by analogy to the standard optical lattice \cite{Bloch2008} approximate
expressions for the band gap 
\begin{align}
BG\approx\sqrt{2JE_{R}^{M}}\label{eqn:ham1gap}
\end{align}
and bandwidth 
\begin{align}
BW\approx16E_{R}^{M}{(J/2E_{R}^{M})^{3/4}e^{-2\sqrt{J/2E_{R}^{M}}}}/{\sqrt{\pi}}\;.
\end{align}
Hence, in this limit $J/E_{R}^{M}\gg1$, the flatness ratio $F$ is
exponentially large. For intermediate values $J/E_{R}^{M}=\mathcal{O}(1)$,
$F$ can be calculated numerically, as shown in Fig.\:\ref{fig:AppFigFlatnessEdge}(a)
(note the logarithmic scale). Even for $J/E_{R}^{M}=1$ the flatness
radio $F\sim300$ is already very large. Further, together with increasing
flatness of the lowest band, also its Berry curvature $\mathcal{F}_{k}^{\alpha}=i\text{Tr}\left\{ P_{k}^{\alpha}\left[(\partial_{k_{x}}P_{k}^{\alpha}),(\partial_{k_{y}}P_{k}^{\alpha})\right]\right\} $
(see also Eq.~(\ref{eqn:Chern})) shows increasing flatness, as indicated
in Fig.\:\ref{fig:AppFigFlatnessEdge}(b).

\begin{figure}[t]
\includegraphics[width=86mm]{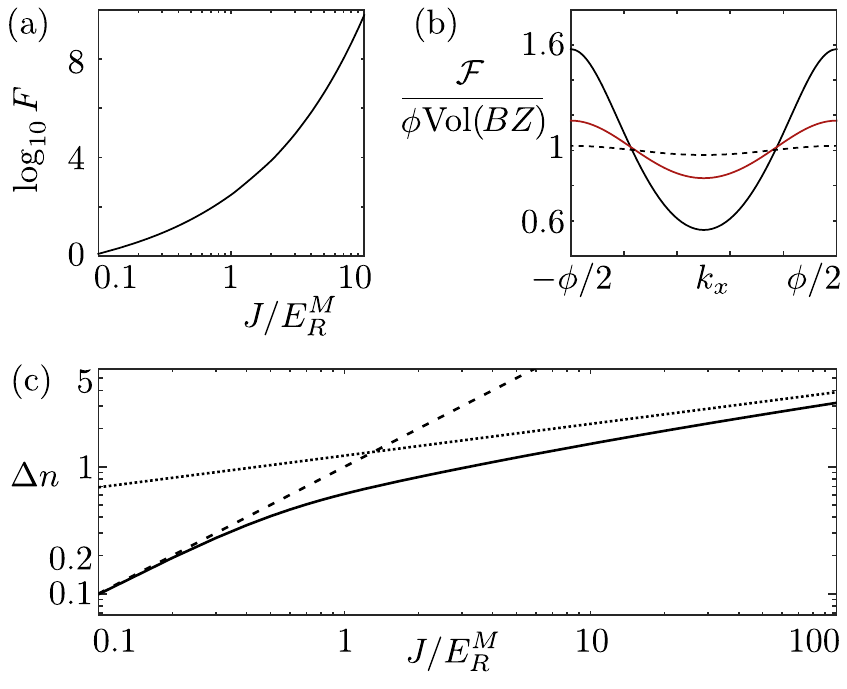}
\caption{(Color online) Illustration of properties of the band structure
of the Hamiltonian \eqref{eqn:ham1}. Panel (a) shows the flatness ratio
$F$ of the band as a function of $\tilde J= J/E_{R}^{M}$. Note the log scale
on the $y$-axis. Panel (b) shows the normalized Berry curvature for
different $J/E_{R}^{M}$. The values of the latter are $0.25$ (black
solid line), $0.5$ (red solid line), $1.0$ (black dotted line). Panel
(c) shows the width of the edge mode wavefunction (for $k_{x}=0$)
defined as $\Delta n=\sqrt{\langle(Y-Y_{\textrm{edge}})^{2}\rangle}/a$.
Note that $\Delta n=0$ indicates perfect localization of the wavefunction
on the edge. The guide lines show asymptotic behavior for $J/E_{R}^{M}\gg1$
(dotted line) and for $J/E_{R}^{M}\ll1$ (dashed line). }
\label{fig:AppFigFlatnessEdge} 
\end{figure}

The form of single particle wavefunctions is given by Eq.\:\eqref{eqn:singleParticleWavefunction}.
The corresponding momentum sector of the Hamiltonian \eqref{eqn:hky}
is 
\begin{align}
H=\sum_{j=1}^{N}E_{j}{\hat{n}}_{j}+J\sum_{j=1}^{N-1}a_{j+1}^{\dagger}a_{j}+H.c.\label{eq:hky_harmonic}
\end{align}
where $E_{j}=\hbar^{2}k_{j}^{2}/2m,k_{j}=k_{x}+\phi(j-j_{0})$. The
$E_{j}$ form a parabolic energy barrier. In particular, edge modes
correspond to the case of $j_{0}\approx1$ or $j_{0}\approx N$. For
large $J/E_{R}^{M}$ the above Hamiltonian is formally identical to
the harmonic oscillator problem $H=\frac{d^{2}}{dy^{2}}+\alpha y^{2}$
on the positive real axis. Mapping it to the discrete case given by
\eqref{eq:hky_harmonic}, for $J/E_{R}^{M}\gg1$ the width of the
edge mode wavefunction is $\langle(Y-Y_{\textrm{edge}})^{2}\rangle_{GS}/a^{2}\approx3/2\sqrt{J/E_{R}^{M}}$.
Thus the width of the edge mode is 
\begin{align}
\Delta n=\sqrt{\langle(Y-Y_{\textrm{edge}})^{2}\rangle}/a=\sqrt{3/2\sqrt{J/E_{R}^{M}}}.\label{eqn:deltaN}
\end{align}
This scaling behaviour with $J/E_{R}^{M}$ is verified numerically
in Fig.\:\ref{fig:AppFigFlatnessEdge}(c) and
indicates that well localized edge modes occur when $J/E_{R}^{M}\lesssim{\cal O}(1)$.

In summary, for $J/E_{R}^{M}\approx1$, the system described by Hamiltonian
\eqref{eqn:ham1} features both (very) flat bands and well-localized
edge modes. Keeping $J$ constant and decreasing $E_{R}^{M}$ has
the effect of exponentially increasing the flatness ratio of the Bloch
bands at the price of algebraic increase of the width of the edge
modes given by Eq.\:\eqref{eqn:deltaN}. Another effect is the decrease
of the gap between the topological bands of Hamiltonian \eqref{eqn:ham1}
as indicated by Eq.\:\eqref{eqn:ham1gap}, which however allows to
utilize the flat band regime.

\begin{figure}[t]
\includegraphics[width=86mm]{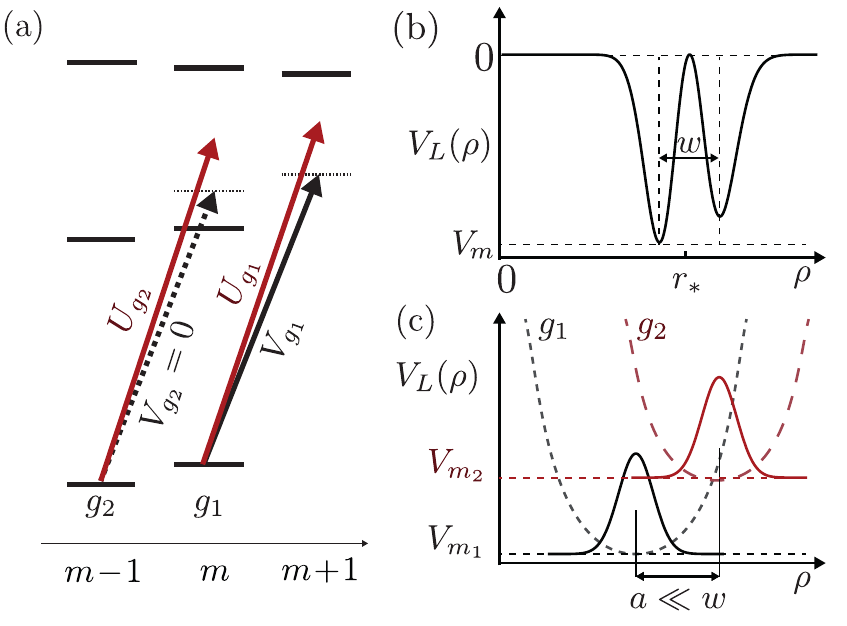} \caption{(Color online) Panel (a) Atomic level scheme for the state-dependent
lattice akin to Ref.\:\cite{Kuhr2011}. Black lines indicate a $\sigma_{+}$-polarized laser tuned to a magic wavelength to make the AC-Stark shift
for atoms in state $g_{2}$ zero. The red laser indicate the second
laser used, and symbols $U_{g_{i}},V_{g_{i}}$ the created optical
potentials (see Appendix \ref{app:spinDependentTorus} for discussion).
Panel (b) Constructing optical potential in the form of parallel rings
using the LG mode with $p=1$ mode results in large ($\sim w$) separation
of subsequent rings in $w\gg\lambda$ limit. Panel (c) Two rings created
by a state-dependent potential. One (black, short dashes) potential
well is for atoms in $g_{1}$ state, other (red, long dashes) for
atoms in $g_{2}$ state. The resulting separation of the ground state
wave functions is $a\sim \lambda \ll w$. }
\label{fig:FigSpinDep} 
\end{figure}

\section{Spin-dependent lattice for the torus geometry}

\label{app:spinDependentTorus} Two coaxial, independent series of
parallel rings coupled at their ends, create the torus geometry as
described in Sec.~\ref{sec:Implementation:Cylindric:torus}. In this Appendix we
detail the possible implementation with state-dependent potentials.

The radial cut through the optical potential created by the
LG beams (for $p=0$), has just a single Gaussian potential well with
width given by the waist $w\gg\lambda/2$ (see Sec.~\ref{sec:Implementation:Cylindric:cylinder}).
Although taking $p=1$ would create two ring-shaped potentials {[}see
Fig.\:\ref{fig:FigSpinDep}(b){]}, their large radial separation
of order $\sim w$ would make Raman coupling of rings of each cylinder
very small.

Instead we create each ring as a state-dependent potential for
atom gas in two stable states $g_{1}$ and $g_{2}$ which are
extremal atomic states of the hyperfine manifold (similar to Ref.\ \cite{Kuhr2011}). Such a gas is then coupled
to the manifold of excited states by a $\sigma_{+}$ polarized laser
light in Laguerre-Gaussian mode $LG_{l,p}$ {[}see Fig.\:\ref{fig:FigSpinDep}(a)
- coupling due to that laser are indicated black{]}. Its wavelength
is chosen as a 'magic' wavelength meaning the AC-Stark shift and the
created optical potential for a $g_{2}$ state is zero $V_{g_{2}}(\mathbf{r})=0$,
and nonzero for the state $g_{1}$: $V_{g_{1}}(\mathbf{r})$. Low
heating rates for state-dependent potentials can be reached for atoms
with large fine-structure splitting such as dysprosium and erbium
\cite{DyEr}. A second laser beam (with a different wavelength, in
\ref{fig:FigSpinDep}(a) marked red) also in a LG mode creates then
optical potentials for both components: $U_{g_{1}}(\mathbf{r})$ and
$U_{g_{2}}(\mathbf{r})$.

The potentials $V_{g_{i}}$ and $U_{g_{i}}(\mathbf{r}),i=1,2$ can
be chosen to have different radii realizing the minimum of the potential
in the radial direction. This allows to microadjust the radial separation
of the minima of felt by each of the components $g_{1}$ and $g_{2}$
on distances $\ll w$ {[}see Fig.\:\ref{fig:FigSpinDep}(c){]}. As
made clear in Appendices \ref{app:MicroscopicDarkStateHamiltonian}
and \ref{app:FloquetTwoCoupledWires} this condition is sufficient
and necessary to be able to reach Raman hopping amplitudes just as
between two neighboring wires of the same cylinder (with a separation
$a=\lambda/2$).

\section{Microscopic derivation of the `dark state' Hamiltonian }

\label{app:MicroscopicDarkStateHamiltonian}

In this Appendix, we present the derivation of the dark state Hamiltonian
$H_{D}$ {[}Eq.\ \eqref{eq: HDarkstate}{]} and the time dependent
potential $H_{RM}(t)$ {[}Eq.\:\eqref{eq: HRaman}{]} resulting from
a far-detuned two-photon Raman transition to an auxiliary level. For
a detailed analysis of the nanoscale dark state potentials, we refer
to Ref.\:\cite{Zoller2016} (see also Ref.\:\cite{Gorshkov2016}).

Assuming a very tight confinement in the $z$-direction, we consider
a single atom moving in the $x$-$y$-plane. Whereas it moves freely
in $x$-direction, we introduce a harmonic confinement potential in
$y$-direction, e.g.\ as a potential well of a standard 1D optical
lattice with depth $V_{y}$ and effective trapping frequency $\omega_{T}\approx\sqrt{4V_{y}E_{R}}$
in harmonic approximation. We obtain 
\begin{align}
H_{\textrm{ext}}(x,y)=-\left(\frac{\hbar^{2}\partial_{x}^{2}}{2m}+\frac{\hbar^{2}\partial_{y}^{2}}{2m}\right)+\frac{1}{2}m\omega_{T}^{2}y^{2}\;.
\end{align}
\begin{figure}
\centering \includegraphics[width=86mm]{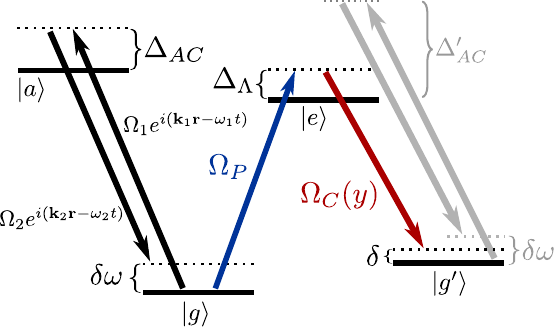}
\caption{(Color online) Level scheme of the atomic $\Lambda$-type configuration formed by
the long lived ground states $\ket{g}$ and $\ket{g'}$ coupled via
Rabi frequencies $\Omega_{P}$ (blue) and $\Omega_{C}(y)$ (red), respectively,
to an excited level $\ket{e}$. Additionally, far detuned Raman beams,
moving waves with amplitudes $\Omega_{1}$ and $\Omega_{2}$, couple
$\ket{g}$ to an auxiliary level $\ket{a}$ (black arrows) and $\ket{g'}$ to $\ket{e}$ (grey arrows).}
\label{fig: levelscheme} 
\end{figure}

Internally, we consider an atomic $\Lambda$-type configuration where
two long lived ground states $\ket{g}$ and $\ket{g'}$ are coupled
to an excited state $\ket{e}$ via a weak probe laser with Rabi frequency
$\Omega_{P}$ (constant in the $x$-$y$-plane) and a strong control
laser, a standing wave with Rabi frequency $\Omega_{C}(y)=\Omega_{C}\sin(k(y-y_{0}))$
(see Fig.\:\ref{fig: levelscheme}). Further, we couple the ground
state $\ket{g}$ to an auxiliary level $\ket{a}$ via two moving waves
with Rabi frequencies $\Omega_{1}\exp(i(\mathbf{k_{1}}\cdot\mathbf{r}-\omega_{1}t))$
and $\Omega_{2}\exp(i(\mathbf{k_{2}}\cdot\mathbf{r}-\omega_{2}t))$
incident in the $x$-$y$-plane. Choosing a large detuning $|\Delta_{AC}|\gg|\Omega_{1,2}|$,
we inevitably introduce a similar off-resonant coupling of $\ket{g'}$
and $\ket{e}$ with possibly slightly different Rabi frequency amplitudes
$\Omega'_{1,2}$ and detuning $|\Delta'_{AC}|\gg|\Omega'_{1,2}|$.
Adiabatic elimination of both transitions leads to AC-Stark shifts
of the ground states $\ket{g}$ and $\ket{g'}$ whose static part
can be absorbed into an appropriate choice of the detunings $\delta=\Omega'-\Omega$,
$\tilde{\Delta}=\Delta_\Lambda+\Omega$. The resulting internal Hamiltonian
of the $\Lambda$-system in a rotating frame is given by 
\begin{align}
 & H_{\textrm{int}}(x,y)=-\tilde{\Delta}\ketbra{e}{e}\nonumber \\
 & \quad+\cos(\delta\mathbf{k}\cdot\mathbf{r}-\delta\omega t)\left(\Omega\ketbra{g}{g}+\Omega'\ketbra{g'}{g'}\right)\nonumber \\
 & \quad+\frac{1}{2}\left(\Omega_{P}\ketbra{e}{g}+\Omega_{C}(y)\ketbra{e}{g'}+\text{h.c.}\right)
\end{align}
where $\Omega=\Omega_{1}\Omega_{2}/(2\Delta_{AC})$ (resp.\ $\Omega'=\Omega'_{1}\Omega'_{2}/(2\Delta'_{AC})$)
denote the two-photon Rabi frequencies and $\delta\omega=\omega_{1}-\omega_{2}$
($\delta\mathbf{k}=\mathbf{k}_{1}-\mathbf{k}_{2}$) the two-photon
detuning. For $\Omega=\Omega'=0$, the $\Lambda$-system possesses
a position dependent dark state $\ket{E_{0}(y)}=\sin\alpha(y)\ket{g}-\cos\alpha(y)\ket{g'}$
where $\tan\alpha(y)=\Omega_{C}(y)/\Omega_{P}$. For simplicity, we
consider in the following the resonant case $\tilde{\Delta}=0$. Then,
for $\Omega_{C},\Omega_{P}\gg\Omega,\Omega'$ the dark state decouples
and we find after projection with $\mathcal{P}(y)=\ketbra{D(y)}{D(y)}$
\begin{align}
\mathcal{P}(y) & H_{\textrm{int}}(x,y)\mathcal{P}(y)=\cos(\delta\mathbf{k}\cdot\mathbf{r}-\delta\omega t)\nonumber \\
 & \cdot\left(\Omega+(\Omega'-\Omega)\cos^{2}\alpha(y)\right)\ketbra{D(y)}{D(y)}\;.\label{eq: darkstateint}
\end{align}
Due to the position dependence of the dark state, the projection of
the kinetic part contained in $\mathcal{P}(y)H_{\textrm{ext}}(x,y)\mathcal{P}(y)$
is non-trivial. Assuming small kinetic energy, $E_{R}/\epsilon^{2},\hbar\omega_{T}\ll\Omega_{P},\Omega_{C}$,
additionally introduced couplings to the bright states can be neglected
\cite{Zoller2016}. However, the dark state is still subject to an
additional conservative potential, the first non-adiabatic correction
\nolinebreak \cite{Zoller2016} 
\begin{align}
V_{na}(y)=\frac{\hbar^{2}}{2m}(\partial_{y}\alpha)^{2}=E_{R}^{2}\frac{\epsilon^{2}\cos^{2}(k(y-y_{0}))}{\left[\epsilon^{2}+\sin^{2}(k(y-y_{0}))\right]^{2}}\;.
\end{align}
This potential represents the nanoscale barrier appearing where $\Omega_{C}(y)\approx0$,
whose height $E_{R}/\epsilon^{2}$ and width $\epsilon/k$ is determined
by the ratio of amplitudes $\epsilon=\Omega_{P}/\Omega_{C}\ll1$ of
the Rabi frequencies constituting the $\Lambda$-system. Physically,
the potential reflects the rapid change of the dark state identity
populating the state $\ket{g'}$ only inside the nanoscale region
of width $\epsilon/k$. Since the second term $\sim\cos^{2}\alpha(y)=|\braket{g'|D(y)}|^{2}$
in equation \eqref{eq: darkstateint} results from the additional
Raman coupling of $\ket{g'}$ to $\ket{e}$, it accordingly contributes
to any relevant hopping matrix element $B_{nn'}$ (equation \eqref{eq: matrixelements})
only to order $\mathcal{O}(\epsilon^{2})$. We neglect it to arrive
at the full dark state Hamiltonian $H(t)=H_{D}+H_{RM}(t)$ [equations
\eqref{eq: HDarkstate} and \eqref{eq: HRaman}].

\section{Floquet analysis of two coupled wires}

\label{app:FloquetTwoCoupledWires}

In this Appendix, we describe the numerical and perturbative calculation
of the Raman assisted tunneling amplitude in the nanoscale double
wire introduced in Sec.~\ref{sec:Implementation:wires}. The
full Hamiltonian of this system $H(t)=H_{D}+H_{RM}(t)$ [equations
\eqref{eq: HDarkstate} and \eqref{eq: HRaman}] is fundamentally
characterized by its discrete time translation symmetry $H(t)=H(t+T)$
with period $T=2\pi/\delta\omega$ and frequency $\delta\omega$.
It describes a periodically driven system for which Floquet's theorem
yields a representation of its time evolution operator 
\begin{align}
U(t,t_{0})=P(t)\;e^{iH_{F}[t_{0}](t-t_{0})}\;.
\end{align}
Here, $P(t)=P(t+T)$ is a unitary, periodic operator with $P(t_{0})=\mathbbm{1}$
and $H_{F}[t_{0}]$ the so called Floquet Hamiltonian. $P(t)$ describes
the micromotion within a single period whereas $H_{F}[t_{0}]$ describes
the effective dynamics after full periods of the driving (starting
at $t_{0}$), the macromotion, in which we are mainly interested in
\cite{Eckardt2016}.

Truncating the Fourier expansion of its eigenstates $H_{F}[t_{0}]$
can be computed numerically. Concretely, we use an expansion up to
frequencies of $\pm25\delta\omega$ and truncate the Hilbert space
of the unperturbed Hamiltonian to the $n=15$ lowest eigenstates to
calculate the effective time-averaged coupling $J\equiv H_{F}^{LR}$
of left and right wire. The result is displayed in Fig.\:\ref{fig: app_coupl}
(solid blue line) and shows clearly the higher order effects bounding
the coupling for large driving amplitude after a linear rise in the
perturbative regime (see also main text). The driving frequency $\delta\omega$
is thereby adjusted self-consistently, such that it compensates the
energy mismatch $\Delta$ between the bare states $\ket{L}$ and $\ket{R}$
and static AC Stark shifts, which are caused by the periodic driving
and hence depends on $\Omega$ and $\delta\omega$, at the maximum
of $J$. 
\begin{figure}
\centering \includegraphics[width=86mm]{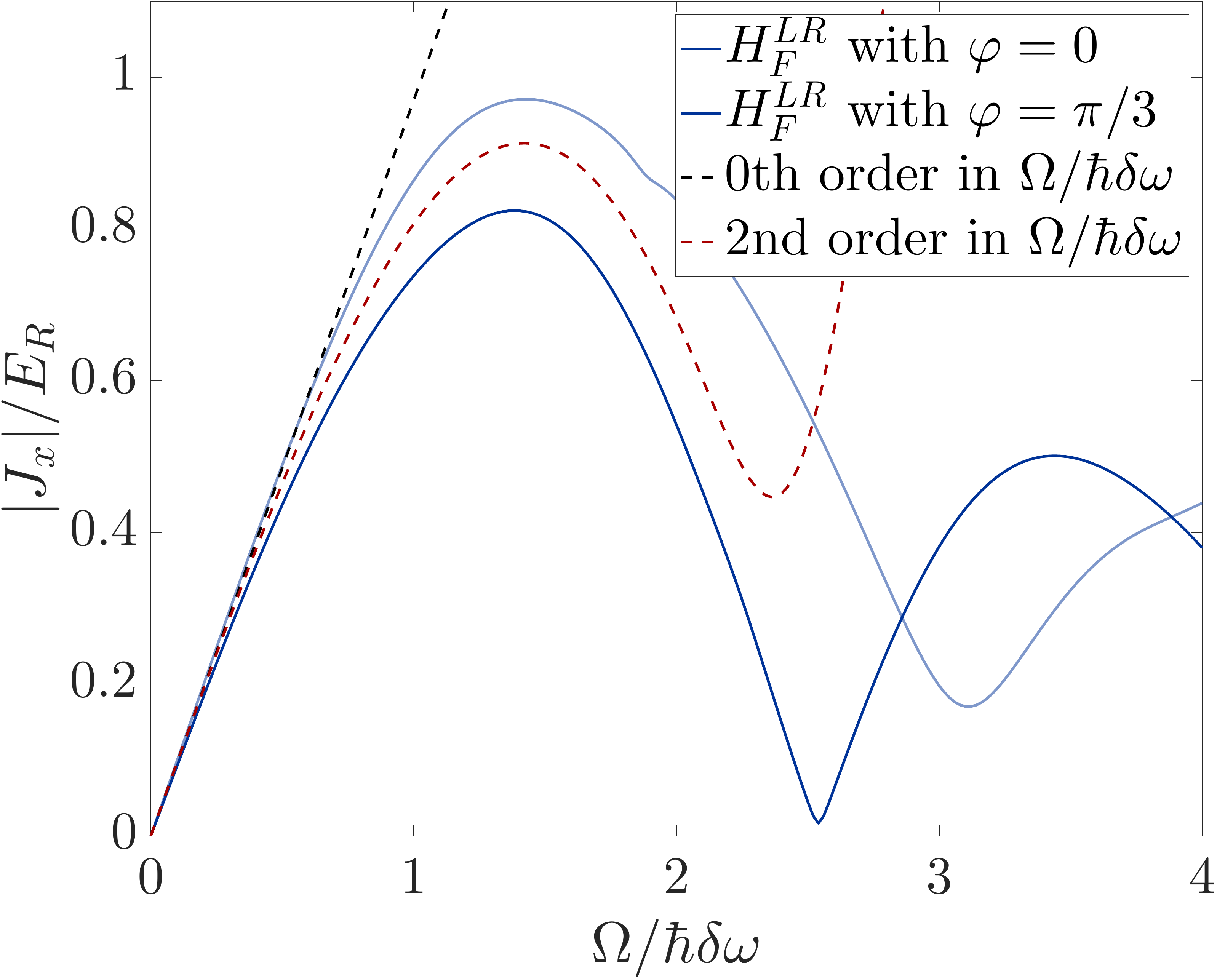}
\caption{(Color online) Raman induced hopping amplitude $J=H_{F}^{LR}$ as a function of the
two photon Rabi frequency $\Omega$ obtained via a numerical Floquet
analysis for two different initial phases (blue lines). Whereas the
strong driving regime shows large phase-dependent effects, the maximal
achievable coupling differs only by a few percent and is well approximated
by the second order expansion (red dashed) of the manifestly phase
independent effective Hamiltonian $H_{\textrm{eff}}$ (see text).
$\epsilon\approx0.1$, $y_{0}\approx0.1a_{H}$, $\delta k_{y}=2k$
and $\delta\omega\approx1.1\Delta/\hbar$ have been chosen to maximize resonant
tunneling. }
\label{fig: app_coupl} 
\end{figure}

So far, we considered a driving of the system starting at definite
time $t_{0}=0$ with vanishing phase at the origin (see equation \eqref{eq: HRaman}).
However, the effect of a periodic driving and therefore $H_{F}[t_{0}]$
typically depend on its phase at initial time \cite{Goldman2014,Goldman2015}.
To quantify how these effect alter the maximal achievable coupling
we include a phase $\varphi_{0}$ to the Hamiltonian, i.e.\ replace
equation \eqref{eq: HRaman} with 
\begin{align}
H_{RM}^{\varphi_{0}}(x,y,t)=\Omega\cos(\delta k_{x}x+\delta k_{y}y-\delta\omega t+\varphi_{0})\;.
\end{align}
As shown in figure \ref{fig: app_coupl}, the coupling, i.e.\ the
Floquet Hamiltonian $H_{F}$, depends strongly on the initial phase
in the strong driving regime. However, the maximal achievable coupling
varies only in the range of a few percent. To quantify this and distill
the time averaged dynamics resulting purely from the periodicity of
the driving, we follow an approach developed in Refs.\ \cite{Goldman2014,Goldman2015,Eckardt2015,Eckardt2016}.

Via a unitary transformation, the phase dependence of $H_{F}$ can
be absorbed into so called kick operators $K(t)$ yielding the representation
of the time evolution operator 
\begin{align}
U(t,t_{0})=e^{-iK(t)}e^{iH_{\text{eff}}(t-t_{0})}e^{iK(t_{0})}
\end{align}
where the effective Hamiltonian $H_{\text{eff}}$ is by definition
constant in time and independent of the phase of the driving at initial
time. It captures only time averaged effects due to the periodicity
of the driving. In contrast, the periodic kick operators $K(t)=K(t+T)$
describe the micromotion within one period. For small driving $\Omega\lesssim\hbar \delta\omega$,
a perturbative expansion in powers of $(\Omega/\hbar \delta\omega)$ is
applicable to compute $H_{\text{eff}}$ and $K$. Using the Fourier
decomposition 
\begin{align}
H(t) & =H_{0}+V(t)\\
V(t) & =\sum_{j=1}^{\infty}V^{(j)}e^{ij\delta\omega t}+V^{(-j)}e^{-ij\delta\omega t}
\end{align}
with $H_{0}=\langle H(t)\rangle_{T=1/\delta\omega}$ one finds for
the effective Hamiltonian \cite{Goldman2014,Eckardt2015}: 
\begin{align}
H_{\textrm{eff}}= & H_{0}+\frac{1}{\delta\omega}\sum_{j=1}^{\infty}\frac{1}{j}\left[V^{(j)},V^{(-j)}\right]\nonumber \\
 & +\frac{1}{2\delta\omega^{2}}\sum_{j=1}^{\infty}\frac{1}{j^{2}}\left(\com{\com{V^{(j)}}{H_{0}}}{V^{(-j)}}+\text{h.c.}\right)\nonumber \\
 & +\frac{1}{3\delta\omega^{2}}\sum_{j,l=1}^{\infty}\frac{1}{jl}\left(\com{V^{(j)}}{\com{V^{(l)}}{V^{(-l-j)}}}\nonumber\right.\\
 & \left.\qquad\qquad-\com{V^{(j)}}{\com{V^{(-l)}}{V^{(l-j)}}}+\text{h.c.}\right)\nonumber \\
 & +\mathcal{O}\left(\frac{1}{\delta\omega^{3}}\right)\;.\label{eq: heff}
\end{align}
The kick operator is given by 
\begin{align}
K(t) & =\sum_{j\neq0}V^{(j)}\frac{1}{i\delta\omega j}e^{ij\delta\omega t}+\mathcal{O}\left(\frac{1}{\delta\omega^{2}}\right)\;.
\end{align}
which is first order in $1/\delta\omega$. Hence, initial time or
phase dependent effects enter only for strong driving, as shown in
figure \ref{fig: app_coupl}. To calculate $H_{\text{eff}}$ we move
into a rotating frame defined by 
\begin{align}
U(t)= & e^{-i/2\left[(\Delta/\hbar-\delta\omega)t+\varphi_{0}\right]}\ketbra{L}{L}\nonumber \\
 & +e^{-i/2\left[(\Delta/\hbar +\delta\omega)t-\varphi_{0}\right]}\ketbra{R}{R}\;.
\end{align}
For our concrete setup, only the first $V^{(j=\pm1)}$ and second
$V^{(j=\pm2)}$ harmonics of the potential are non-vanishing. We find
the effective Hamilton $H_{\textrm{eff}}$ to second order in $\Omega/\hbar \delta\omega$
in the basis spanned by wave functions in left and right wire $(\Psi_{L}(x),\Psi_{R}(x))$
to be 
\begin{align}
H_{\textrm{eff}}&(x)=  \sum_{n=L,R}\left[\frac{\hbar^2}{2m}\left(-i\partial_{x}+\frac{\Omega^{2}}{(\hbar \delta\omega)^{2}}A_{n}^{(2)}\right)^{2}\right.\nonumber \\
 & \left.\vphantom{\frac{1}{2m}\left(-i\partial_{x}+\frac{\Omega^{2}}{(\hbar \delta\omega)^{2}}A_{n}^{(2)}\right)^{2}}+\left(\frac{\sigma_{n}}{2}(\hbar \delta\omega-\Delta)+\frac{\Omega}{\hbar \delta\omega}\;\epsilon_{n}^{(1)}+\frac{\Omega^{2}}{(\hbar \delta\omega)^{2}}\epsilon_{n}^{(2)}\right)\right]\ket{n}\bra{n}\nonumber \\
 & +e^{-i\delta k_{x}x}\left[\kappa^{(0)}+\frac{\Omega^{2}}{(\hbar \delta\omega)^{2}}\kappa^{(2)}\right]\ket{L}\bra{R}+\text{h.c.}\nonumber \\
 & +\mathcal{O}\left(\frac{\Omega^{4}}{(\hbar\delta\omega)^{3}}\right)\label{eq: heffpert}
\end{align}
where $\sigma_{L}\equiv1$ and $\sigma_{R}\equiv-1$. With the hopping
matrix elements $B_{nn'}$ 
\begin{align}
B_{nn'}\equiv\braket{n|e^{-i\delta k_{x}\hat{y}}|n'}\;,\label{eq: matrixelements}
\end{align}
we have obtained a constant second order gauge potential 
\begin{align}
\frac{\Omega^{2}}{(\hbar\delta\omega)^{2}}A_{n}^{(2)}=\sigma_{n}\frac{\Omega^{2}}{(\hbar \delta\omega)^{2}}\frac{|B_{LR}|^{2}}{32}\delta k_{y}\;,
\end{align}
AC stark shifts of the individual levels 
\begin{align}
\frac{\Omega}{\hbar\delta\omega}\epsilon_{n}^{(1)} & =-\sigma_{n}\frac{\Omega^{2}}{\hbar\delta\omega}\frac{|B_{LR}|^{2}}{8}\;,\nonumber \\
\frac{\Omega^{2}}{(\hbar\delta\omega)^{2}}\epsilon_{n}^{(2)} & =\frac{\Omega^{2}}{(\hbar\delta\omega)^{2}}\left[\left(\frac{|B_{nn}|^{2}}{4}+\frac{1}{32}|B_{LR}|^{2}\right)\frac{\hbar^2\delta k_{y}^{2}}{2m}\right.\nonumber \\
 & \qquad\qquad-\left.\sigma_{n}\frac{|B_{LR}|^{2}}{16}(\hbar\delta\omega-\Delta)\right]\;,
\end{align}
and the Raman induced couplings 
\begin{align}
\kappa^{(0)} & =\Omega\;\frac{B_{LR}}{2}\;,\nonumber \\
\frac{\Omega^{2}}{(\hbar\delta\omega)^{2}}\kappa^{(2)} & =-\frac{\Omega^{3}}{(\hbar\delta\omega)^{2}}\left(\frac{B_{LR}}{8}|B_{LL}-B_{RR}|^{2}\right.\label{eq: coupling2nd}\\
 & \qquad\left.-\frac{B_{10}^{*}}{16}\left(B_{LL}-B_{RR}\right)^{2}+\frac{B_{LR}}{32}|B_{LR}|^{2}\right)\;.\nonumber 
\end{align}
The absolute value of the coupling $J=H_{\textrm{eff}}^{01}=e^{-i\delta k_{x}x}\left[\kappa^{(0)}+\frac{\Omega^{2}}{(\hbar\delta\omega)^{2}}\kappa^{(2)}\right]$
is shown to zeroth and second order in $\Omega/\hbar \delta\omega$ in Fig.\:\ref{fig: app_coupl}.
The bounding effect of the higher order processes which involve on-site
and multiple intersite hopping processes is clearly visible. The driving
frequency has been chosen to obtain resonant tunneling at the maximal
amplitude, i.e.\ such the  Stark shifts $\epsilon_{1}$ and $\epsilon_{2}$
are canceled when $\Omega$ is chosen to maximize $J$. Note that
the initial phase $\varphi_{0}$ does not enter the effective Hamiltonian
$H_{\text{eff}}$. In contrast, the kick operator depends explicitly
on $\varphi_{0}$ and is given by 
\begin{align}
K(t,x)= & \sum_{n=L,R}\frac{\Omega}{2i\hbar \delta\omega}e^{i(\delta\omega t-\delta k_{x}x-\varphi_{0})}B_{nn}\ketbra{n}{n}+\text{h.c.}\nonumber \\
 & +\frac{\Omega}{4i\hbar \delta\omega}e^{i(2\delta\omega t-\delta k_{x}x-2\varphi_{0})}B_{RL}\ketbra{R}{L}+\text{h.c.}\nonumber \\
 & +\mathcal{O}\left(\frac{\Omega^{2}}{(\hbar\delta\omega)^{2}}\right)\;.
\end{align}
It leads to phase dependent effects of order $\Omega/\hbar \delta\omega$.
To conclude, we compare our result \eqref{eq: heffpert} to an approach
presented in Ref.\:\cite{Ketterle2013}. Here, a non-perturbative
resummation of on-site hopping processes mediated by the matrix elements
$B_{nn}$ was performed via a transformation to another rotating frame.
Time-averaging in this frame yields the well-known Bessel function
renormalization of the coupling, keeps however only terms linear in
$B_{LR}$. In our case, all matrix elements $B_{nn'}$ are of order
unity. Hence, a selective resummation of on-site processes does not
improve the agreement to the non-perturbative numerical analysis.
On the contrary, the inclusion of the last term in \eqref{eq: coupling2nd},
neglected in Ref.\:\cite{Ketterle2013}, leads in our case to a better
agreement of the second order perturbation theory than the partially resummed result of Ref.\:\nolinebreak
\cite{Ketterle2013}.

\bibliographystyle{apsrev}

\begin{thebibliography}{10}
\bibitem{note4}{A magnetic length $l_B \gg \lambda/2$ could be obtained with a high-order multiphoton Raman process.}
\end{thebibliography}

\end{document}